\documentclass[twocolumn]{aastex63}

\newcommand{\cii}{[\ion{C}{2}] }
\newcommand{\Cii}{[\ion{C}{2}]}

\received{Dec 8, 2020}
\revised{Jan 20, 2021}
\accepted{Jan 25, 2021}
\submitjournal{ApJ}

\shorttitle{\cii And CO Emission Along the Bar and Counter-Arm of NGC~7479}
\shortauthors{Fadda, Laine, \& Appleton}
\graphicspath{{./}{figures/}}
\begin{document}

\title{\cii and CO Emission Along the Bar and Counter-Arms of NGC~7479\footnote{Based on SOFIA observations with FIFI-LS.}}

\correspondingauthor{Dario Fadda}
\email{darioflute@gmail.com, dfadda@sofia.usra.edu}

\author[0000-0002-3698-7076]{Dario Fadda}
\affiliation{SOFIA Science Center, USRA, NASA Ames Research Center, M.S. N232-12 Moffett Field, CA 94035, USA}
\author[0000-0003-1250-8314]{Seppo Laine}
\affiliation{IPAC, Mail Code 314-6, Caltech, 1200 E. California Blvd., Pasadena, CA 91125, USA}
\author[0000-0002-7607-8766]{Philip N. Appleton}
\affiliation{IPAC, Mail Code 314-6, Caltech, 1200 E. California Blvd., Pasadena, CA 91125, USA}



\begin{abstract}
We present new {\it SOFIA} [\ion{C}{2}] and {\it ALMA} CO$_{J = 1 \rightarrow 0}$ observations of the nearby asymmetric barred spiral galaxy NGC~7479. The data, which cover the whole bar of the galaxy and the counter-arms visible in the radio continuum, are analyzed in conjunction with a wealth of existing visible, infrared, radio, and X-ray data.  As in most normal galaxies, the [\ion{C}{2}] emission is generally consistent with emission from cooling gas excited by photoelectric heating in photo-dissociation regions. However, anomalously high [\ion{C}{2}]/CO ratios are seen at the two ends of the counter-arms. Both ends show shell-like structures, possibly bubbles, in H$\alpha$ emission. In addition, the southern end has [\ion{C}{2}] to infrared emission ratios inconsistent with normal star formation. Because there is little \ion{H}{1} emission at this location, the [\ion{C}{2}] emission probably originates in warm shocked molecular gas heated by the interaction of the radio jet forming the counter-arms with the interstellar medium in the galaxy. At two other locations, the high [\ion{C}{2}]/CO ratios provide evidence for the existence of patches of CO-dark molecular gas. The [\ion{C}{2}] and CO observations also reveal resolved velocity components along the bar. In particular, the CO emission can be separated into two components associated to gas along the leading edge of the bar and gas trailing the bar. The trailing gas component that amounts to approximately 40\% of the gas around the bar region may be related to a minor merger.
\end{abstract}



\section{Introduction} \label{sec:intro}

Bars are common features in spiral galaxies. Recent infrared studies estimate that in the local Universe approximately 70\% of the spiral galaxies have bars~\citep[see, e.g.,][]{Eskridge2000}. Since this percentage declines at higher redshifts, the presence of a bar has been seen as a sign of galaxies reaching full maturity after several episodes of merging~\citep{Kartik2008}.
Since most galaxies have companions \citep{Zaritsky1997}, minor mergers (with galaxy masses 10 to 20 times smaller than the main galaxy) are likely common events during the life of a galaxy \citep[see, e.g.,][]{Jogee2009}.  Such mergers can play a major role in triggering a stellar bar that efficiently channels gas into the nucleus \citep{Mihos1994}, and have been proposed as a mechanism for the formation of active galactic nuclei \citep[AGNs, see e.g.,][]{taniguchi1999, Kendall2003, Kaviraj2014}. In a merger, the loss of angular momentum in the gaseous component of the interstellar medium (ISM) leads to an inflow of gas toward the galactic nucleus \citep{Barnes1991,Blumenthal2018} causing the growth of the central black hole. Once the central black hole reaches a critical mass \citep{Ishibashi2012}, the AGN starts energizing the surrounding medium via winds, jets, and radiation. This triggers star formation in the surrounding gas and suppresses further gas inflow by blowing the gas out, a process generally known as AGN feedback~\citep[see, e.g.,][]{silk2013}. 

\begin{deluxetable*}{cccccccc}[!t]
\tablecolumns{8}
\tablewidth{0pt}
\tablecaption{ Log of FIFI-LS observations \label{tab:log}}
\tablehead{
\colhead{Observation}\vspace{-0.2cm} &\colhead{Flight}& \colhead{AOR} &\colhead{Starting} &
\colhead{Exposure}&\colhead{Barometric} &\colhead{Zenithal} & \colhead{Zenithal} \\
\colhead{Date}\vspace{-0.2cm}&\colhead{Number}& \colhead{ID} &\colhead{Time} & \colhead{Time}& \colhead{Altitude} & \colhead{Angle} & \colhead{Water vapor}\\ 
\colhead{[UT]} & \colhead{} & \colhead{} & \colhead{[UT]} & \colhead{[minutes]} &\colhead{[feet]} & \colhead{[degs]} & \colhead{[$\mu$m]}
}
\startdata
         2019 05 14 & 570 & 07\_0154\_6 & 09:57:25 & 38 & 42000 & 68.4 -- 61.7 & 3.0 [start]\\
         2019 05 14 & 570 & 07\_0154\_7 & 10:36:19 & 38 & 43000 &
         61.2 -- 54.0 & 3.1 [start] -- 3.2 [end]\\
         2019 05 15 & 571 &  07\_0154\_6 &10:25:12 & 35 & 43000 & 65.9 -- 59.4 &3.2 [start]\\
         2019 05 15 & 571 &  07\_0154\_7 &11:01:39 & 36 & 43000 & 59.3 -- 52.1 &3.7 [end]
\enddata
%\vspace{-0.8cm}
\tablecomments{Observations on the same date were made during one flight leg. Water vapor measurements were taken at the beginning and end of each observation, except for 2019 May 14 when a change of altitude occurred between the AORs, and the water vapor measurement was done after the altitude change. The zenithal angle varied linearly during the observations between the two reported values.}
\end{deluxetable*}

Several observations support the hypothesis that bars act as channels to drive gas from the arms to the nucleus. Barred galaxies have shallower metallicity gradients than unbarred ones \citep{Martin1994}, suggesting that gas is radially transported along the bar. Enhanced star formation is common in the central regions of many barred galaxies~\citep{Ho1997} but not in all of them. An anti-correlation between the presence of a bar and the central atomic gas content could be interpreted as depletion of gas in barred galaxies, possibly resulting from enhanced star formation~\citep{Laine1998, Masters2012}. Although the correlation between bars and AGNs is not clear, there is a definitive correlation between the gaseous absorbing column density towards type~2 Seyfert nuclei and the presence of stellar bars \citep{Maiolino1999}, suggesting that bars are effective in driving gas inward to enshroud galactic nuclei. Finally, molecular gas along bars has been directly observed through CO observations. \citet{Sakamoto1999} and \citet{Kartik2005} found that barred spirals have higher molecular gas concentrations in the central kiloparsec than unbarred galaxies, which is consistent with radial inflow driven by bars.

% Figure with coverage map
\begin{figure}
\begin{center}
\includegraphics[width=0.48\textwidth]{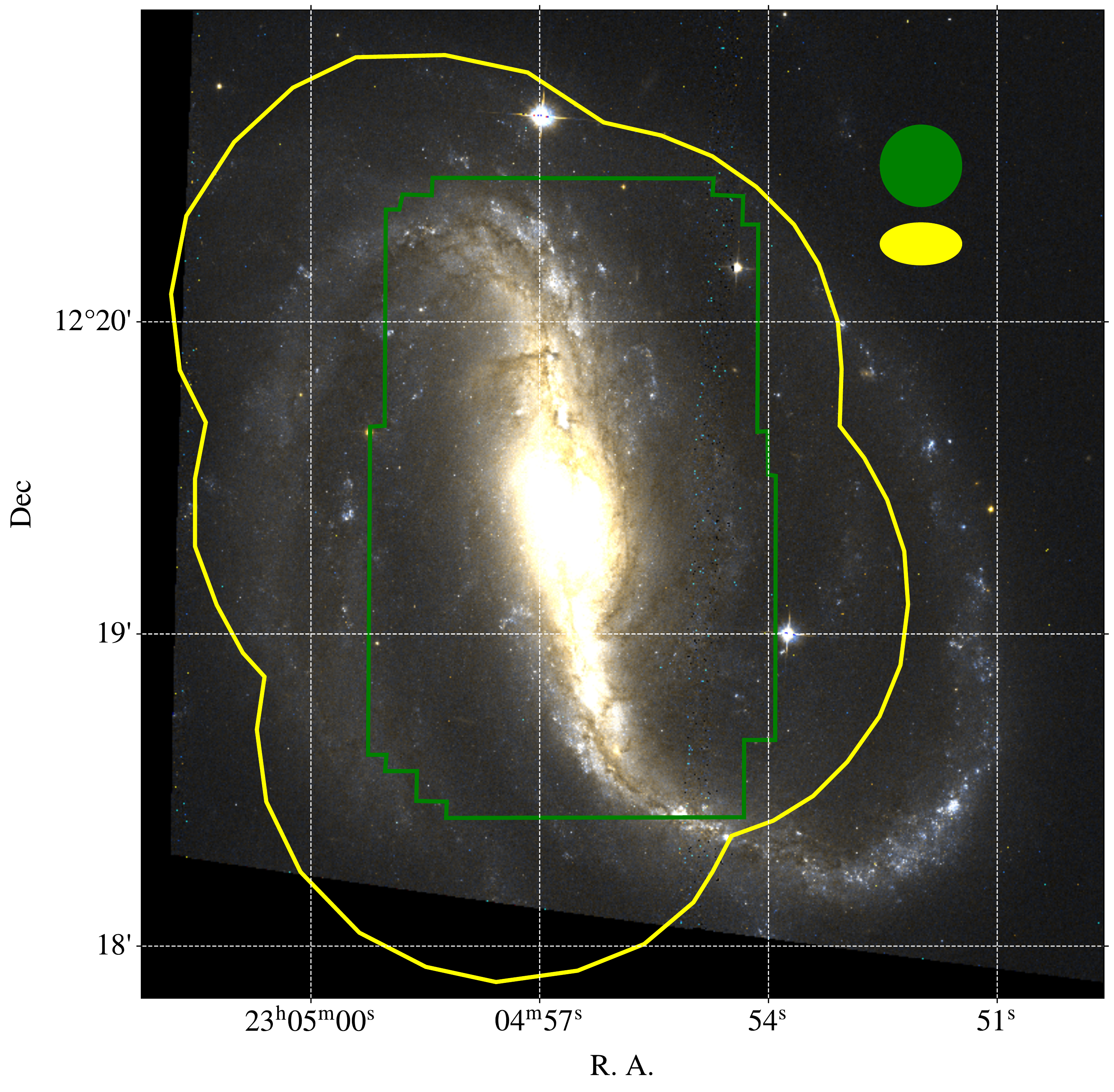}
\end{center}
\caption{Coverage of the FIFI-LS and ALMA observations over a two-color visible image from {\it HST} archival observations (combination of bands F555W and F814W). The green contour defines the region covered by FIFI-LS with at least 500~seconds of on-source integration. The filled green circle corresponds to the beam of FIFI-LS at 159~$\mu$m, the redshifted wavelength of [\ion{C}{2}] emission from NGC~7479. The yellow contour shows the coverage of the ALMA CO observations. The yellow ellipse corresponds to the beam of the ALMA CO observations in the configuration used.
}
\label{fig:coverage}
\end{figure}

Observing molecular gas is an excellent way to study the gas inflow along bars. Because of the depletion due to enhanced star formation or a phase transition into molecular gas, neutral atomic hydrogen along bars can be hard to detect or non-existent~\citep{Masters2012, Laine1998}. Moreover, since the H$\alpha$ emission in the ISM arises from diffuse ionized gas or in HII regions, it will be necessarily biased by regions of active star formation, and is subject to extinction by dust. 
Many studies have traced molecular gas in galaxy bars through CO emission \citep[see, e.g.,][]{Sakamoto1999,1999ApJ...511..709L,Regan1999,Kartik2005}. 

Another excellent tracer of diffuse gas that is not significantly affected by extinction in the ISM is the [\ion{C}{2}] far-IR line at 157.741~$\mu$m. Although observed in many galaxies with {\it Herschel} \citep[see, e.g.,][]{Herrera2015} and SOFIA \citep[see, e.g.,][]{Pineda2018,Bigiel2020}, it has never been used for detailed studies of galaxy bars. The only published study with {\em Herschel} data is limited to the circumnuclear region of the barred spiral NGC~1097~\citep{Beirao2012}. 

[\ion{C}{2}] emission can complement CO studies of the distribution and state of the molecular gas because it most commonly arises in photodissociation regions (PDRs) surrounding star formation locations. In most observations of nearby galaxies [\ion{C}{2}]  generally traces a mix of warm molecular gas heated by the photoelectric ejection of electrons from polycyclic aromatic hydrocarbons (PAH) and small grains, and ionized gas \citep{Draine1978,Tielens1985,Bakes1998}.
For this reason [\ion{C}{2}] emission is commonly used in normal galaxies as a star formation rate indicator \citep[see, e.g.][]{Stacey1991,Malhotra2001,DeLooze2011,Diaz-Santos2014,Herrera2015}. Care must be taken to blindly use [\ion{C}{2}] observations as a proxy for star formation. For example, in addition to correcting for the fraction of ionized gas, this line is sensitive to purely neutral atomic gas \citep{Croxall2017}. It can also reveal the presence of molecular gas in regions of low metallicity that are usually CO-dark \citep{Wolfire2010,Jameson2018,Madden2020,Chevance2020} and may form a significant fraction of the diffuse ISM in our own Galaxy \citep{Pineda2013}. 

In addition, warm molecular gas heated by turbulence and shocks has been shown to emit significant amounts of [\ion{C}{2}] in areas devoid of significant star formation, but in regions with diffuse UV radiation and very strong mid-IR pure-rotational signatures of warm H$_2$ \citep{Appleton2013,Peterson2018}. Such observations of shock-heated intergalactic warm H$_2$ also exhibit very broad [\ion{C}{2}] line widths (400--600 km s$^{-1}$) and unusually high [\ion{C}{2}]/FIR and [\ion{C}{2}]/PAH ratios. Models of warm molecular gas shocks \citep{Appleton2017} in Stephan's Quintet show that low velocity magnetic shocks are likely responsible for the very strong H$_2$ and [\ion{C}{2}] emission.  Recently, further evidence of shock-enhanced [\ion{C}{2}] emission in a different environment was discovered near the ends of a radio jet in NGC~4258, where large quantities of warm H$_2$ were detected \citep{Appleton2018}. The gas was found to correlate not only with warm mid-IR H$_2$ emission, but also with soft X-ray emission relating to the activity of the jet in the inner regions of the galaxy. These results are very relevant to this paper, since NGC~7479, like NGC~4258, also shows a large-scale radio jet that may be interacting with its own ISM. 

We present the analysis of new [\ion{C}{2}] and CO observations of the nearby strongly barred galaxy~NGC~7479 ($cz=2381$~km~s$^{-1}$). The galaxy shows a clear signature of minor merger, visible along the bar of the galaxy~\citep{Quillen1995,Laine1999,Martin2000}, and hosts an AGN.  NGC~7479 also exhibits an S-shaped 10-kpc scale radio continuum structure emanating from the nucleus. These counter-arms, likely caused by a radio-jet originating from the nucleus, were discovered by \citet{Laine2008} and have polarization vectors aligned along the main ridge-line of the structure. We present the first {X-ray detection} of this jet-like structure using archival Chandra data. The radio continuum structure is remarkably similar to the ghostly counter-arms in the nearby galaxy NGC~4258 \citep{Appleton2018}. In NGC~4258, about 40\% of the [\ion{C}{2}] emission in the central region comes from molecular gas excited by shocks and turbulence due to the jet propagating near the plane of the disk. The jet collides with dense clumps of gas in the thick disk and changes direction and dissipates its energy over a wide area of the galaxy. A similar scenario was invoked by \citet{Laine2008} to explain the jet-like structure in NGC~7479. 

Here we present intriguing evidence of [\ion{C}{2}] emission that is spatially coincident with the jet emission, as well as [\ion{C}{2}] emission emanating from the speculated location of the merged companion about 17\arcsec\ north of the nucleus, and elsewhere along the bar.

Throughout this paper, we use $H_0$~=~70\, km\, s$^{-1}$ \, Mpc$^{-1}$, $\Omega_{\rm m}$ = 0.3, and $\Omega_{\rm \Lambda}$ = 0.7. We also refer
to the ground rotational transition $J = 1 \rightarrow 0$  of the most common $^{12}$C$^{16}$O isotopologue as simply CO.

\section{Observations and Data} \label{sec:observations}
\subsection{SOFIA Observations}

The new SOFIA observations were part of the SOFIA Cycle 7 observing program 07\_0154. The Field Imaging Far-Infrared Line Spectrometer \citep[FIFI-LS,][]{2018JAI.....740003F, 2018JAI.....740004C} was used to map the [\ion{C}{2}]~157.741~$\mu$m (rest frame) line. For NGC~7479 that line corresponds to the observer frame wavelength of 159~$\mu$m. The spectral resolving power of FIFI-LS is 1167, meaning that an unresolved line has a FWHM of 257~km~s$^{-1}$. The spatial resolution of the instrument is 15.6~arcseconds, corresponding to 2.5~kpc at the distance (34.2 Mpc) of our galaxy. FIFI-LS is a dual channel instrument. Parallel observations were obtained at 88.3~$\mu$m. Unfortunately, those observations had an insufficient signal-to-noise ratio to derive any science results, and consequently they are not discussed in this paper.

The data were acquired in two consecutive flights (2019 May 14 and 15) for a total of approximately 2.5~hr of flight time (see Table~\ref{tab:log}). For each flight two Astronomical Observation Requests (AORs) were observed during the same leg, one AOR to cover the northern and another AOR to cover the southern part of the bar. The two flight legs were almost identical, and the observations were acquired in very similar atmospheric conditions, resulting in a homogeneous data set. 

Figure~\ref{fig:coverage} shows the extent of the region covered on top of a visible image and the exposure map of ALMA CO observations. The observations were performed in the chop--nod mode with the secondary mirror chopping between the galaxy and reference fields on the two sides of the galaxy, each at a 200~arcsecond distance from the center of the galaxy. Since the instantaneous field of view of FIFI-LS in the red array is approximately $60\arcsec\times60\arcsec$, we covered the total mapped field with two pointings.

Some dithering was performed to reduce the effect of bad pixels and to improve the recovery of the point spread function (PSF) in the images, since the size of the spatial pixel of FIFI-LS (12\arcsec) is not small enough to recover the shape of the PSF. The data were reduced using the FIFI-LS pipeline \citep{2020ASPC..Vacca}. In particular, the data were corrected for atmospheric transmission using the ATRAN model \citep{1992nstc.rept.....L}, and the values of the zenithal water vapor burden were estimated during the observations.

The reduced data were projected into spectral cubes with a fine grid of 3\arcsec\ sampling using a Gaussian spectral kernel with a dispersion equal to 1/4 of the spectral resolution  and a Gaussian spatial kernel with a dispersion equal to 1/2 the spatial resolution. These parameters produced a data cube that conserves the instrumental spectral and spatial resolutions.

% Figure with 24um observations
\begin{figure}
\begin{center}
\includegraphics[width=0.45\textwidth]{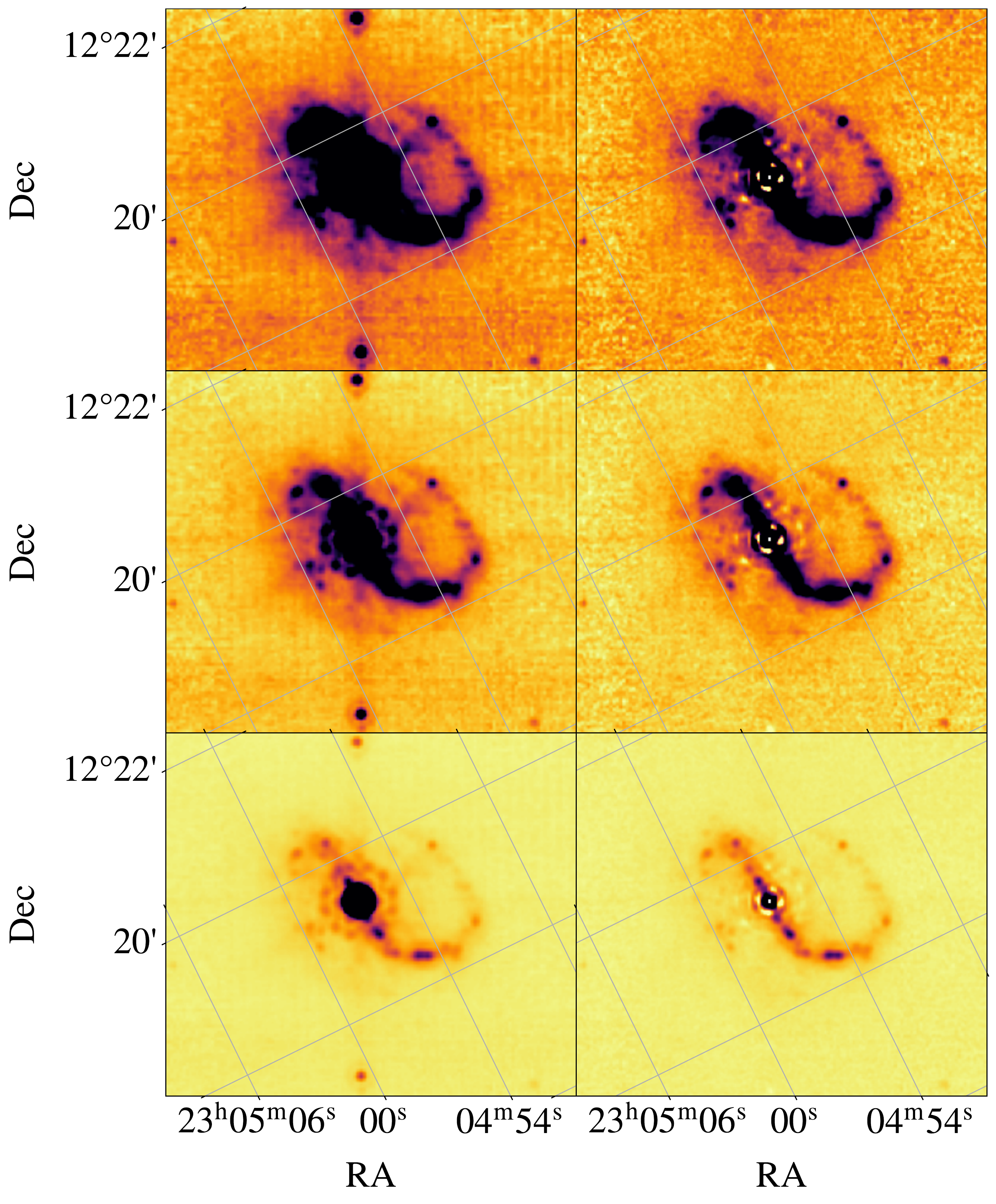}
\end{center}
\caption{Comparison between the archived MIPS 24~$\mu$m observations (left) and our reduction after subtraction of the nuclear PSF (right), shown with three different top brightness cuts (2, 4, and 20 MJy/sr). Artefacts such as residual "jail bars," regularly alternating columns with higher fluxes, and ghost sources, due to memory effects and the dithering pattern of the observation, are visible in the top-left panel. The different brightness cut levels show how the wings of the PSF of the nucleus affect the flux measurements in different parts of the galaxy. The image is oriented according to the MIPS array direction to better show the instrumental features.
}
\label{fig:mips24}
\end{figure}

\subsection{Spitzer Observations}
The galaxy has been observed with the IRAC \citep{2004ApJS..154...10F} and MIPS \citep{2004ApJS..154...25R} instruments onboard the {\it Spitzer Space Telescope} \citep{2004ApJS..154....1W}. We retrieved the relevant data from the {\it Spitzer} Heritage Archive, and found that the IRAC archival data were directly usable, while the MIPS 24~$\mu$m image still contained artefacts, and was dominated by the point spread function (PSF) of the bright nucleus of the galaxy. We therefore produced another mosaic starting from the basic calibrated data (BCDs). 
\begin{figure*}[t!]
\begin{center}
\includegraphics[width=0.7\textwidth]{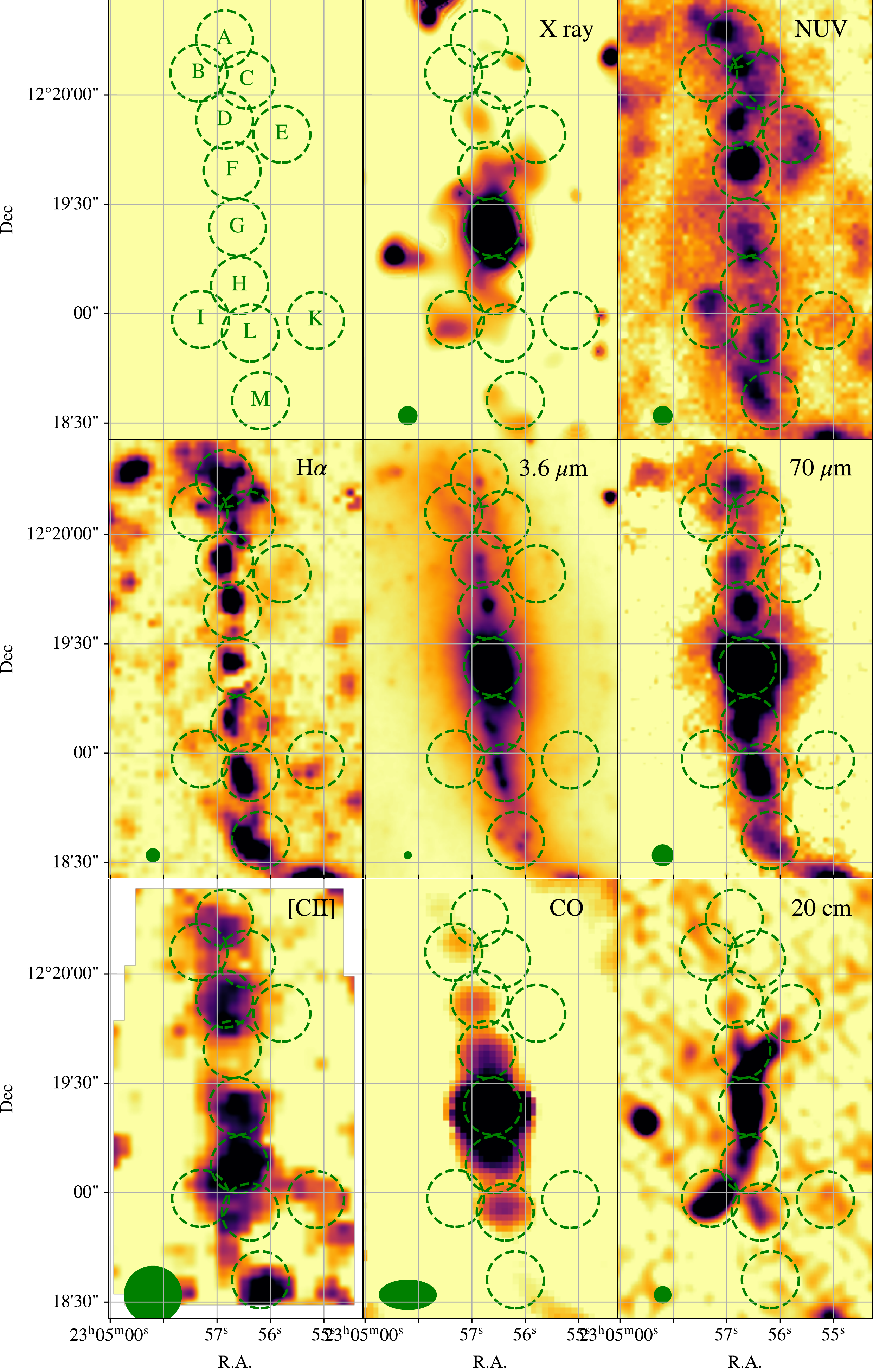}
\end{center}
\caption{Multiwavelength panoramic of NGC~7479's bar. The selected apertures which cover interesting parts of the bar with a diameter equal to the spatial resolution of FIFI-LS are marked in the different images. The green circle at the lower left corner of each panel shows the spatial resolution of the corresponding observation.
}
\label{fig:multiwave}
\end{figure*}
In particular, we removed a pattern of "jail bars," a variation in brightness that repeats every four columns in the BCDs, and is due to the reading mode of the detector. The pattern is visible in the top-left panel of Figure~\ref{fig:mips24}, as well as in a residual gradient in the background. To remove these artefacts, we coadded every fourth column in each BCD and fitted a third-degree Chebyshev polynomial. This average pattern was then subtracted from the respective columns in the original BCDs.

Other visible artefacts in Figure~\ref{fig:mips24} include two symmetric sources at the top and bottom of the image.  These ghost sources are due to latencies in the detector response. Two other ghosts are not visible in the combined image since they fall close to the nucleus of the galaxy. To remove these spurious sources we subtracted the previous four BCDs from each BCD, scaled as in \citet[][section 4.5]{Fadda2006}.

Finally, to remove the wings of the PSF of the bright central source, we used STinyTim\footnote{ \url{https://irsa.ipac.caltech.edu/data/SPITZER/docs/dataanalysistools/tools/contributed/general/stinytim/}} to produce synthetic PSFs. For each BCD, we generated a PSF at the position of the nucleus in the BCD with an oversampling factor of ten. Since this observation has been obtained in the "compact source" mode (see chapter 3.1.1 of the MIPS handbook\footnote{\url{https://irsa.ipac.caltech.edu/data/SPITZER/docs/mips/mipsinstrumenthandbook/}}), we used the predicted positions on the sky recorded in the header as CSM\_SKY as inputs for STinyTim for the displacement of the focal plane along the scan direction. We computed a stack of 100 point source realizations (PSR) by integrating the synthetic PSF in MIPS pixels centered at 100~different positions in the brightest central pixel of the galactic nucleus. To find the best approximation, we maximized the cross-correlation between each BCD and the PSRs. Finally, we minimized the sum of the squares of the differences between each BCD and the optimal PSR to compute the normalization factor. After subtracting these PSRs from all the BCDs, we made a new mosaic using the MOPEX software \citep{2005PASP..117.1113M}. 
As shown in Figure~\ref{fig:mips24}, there are parts of the first and second Airy rings which are slightly over-subtracted. This is due to the limitations of the STinyTim model. The fine details of the PSF depend in fact on the parameters of the optical system that are based on the design rather than the performance of the instrument. On the other hand, empirical PSFs work well only if derived from many point sources in the same observation, such as are available in a wide field survey. Experiments with empirical PSFs from other MIPS observations did not yield a better subtraction. 

Nevertheless in NGC~7479, the subtraction of the synthetic PSF made it possible to obtain more accurate flux density measurements in the parts of the galaxy covered by our new FIFI-LS SOFIA observations.
To give an idea of how much flux is contained in the artefacts and wings of the PSF, we measured the intensity of the ghost sources at the bottom and top of the image.  The ghost fluxes are 0.65\% of that of the central source. These ghost sources, because of the dithering pattern of the observations, appear also in the central part of the galaxy. The first Airy ring contains 30\% of the total flux, while the knots in the secondary Airy ring have 10\% of the total flux. Without removing artefacts and subtracting the PSF of the central source, measurements of the 24~$\mu$m flux along the bar would be severely biased.

\begin{deluxetable*}{cccccccccc}[!t]
\tablecolumns{10}
\tablewidth{0pt}
\tablecaption{Properties of galaxy regions \label{tab:regions}}
\tablehead{
\colhead{Region} &
\colhead{Center} & 
\colhead{L(FIR)} &
\colhead{L([\ion{C}{2}])} &
\colhead{L(CO)} &
\colhead{L(PAH)} &
\colhead{Z} &
\colhead{T$_{ISM}$} &
\colhead{$M_*$} &
\colhead{sSFR} \\
\colhead{Label} &
\colhead{[J2000]} &
\colhead{[$10^9\, L_\odot$]} &
\colhead{[$10^6\, L_\odot$]} &
\colhead{[$10^3\, L_\odot$]} &
\colhead{[$10^8\, L_\odot$]} &
\colhead{[Z$_\odot$]} &
\colhead{[K]} &
\colhead{[$10^9\, M_\odot$]} &
\colhead{[$10^{-10}\, yr^{-1}$]}
}
\startdata
A & 23:04:56.89 +12:20:14.9 & 0.87& 6.19 $\pm$ 0.87& 1.52 $\pm$ 0.11& 0.93&0.8 & 22.1$^{+1.6}_{-1.2}$& 0.88$^{+0.14}_{-0.01}$& 1.86$^{+0.36}_{-0.03}$\\  
B & 23:04:57.37 +12:20:05.5 & 0.90& 4.70 $\pm$ 0.99& 2.50 $\pm$ 0.16& 0.89&1.1 & 21.9$^{+2.4}_{-0.4}$& 1.08$^{+0.01}_{-0.85}$& 0.52$^{+0.50}_{-0.01}$\\  
C & 23:04:56.48 +12:20:03.6 & 0.90& 6.30 $\pm$ 0.98& 0.75 $\pm$ 0.03& 0.91&0.8 & 21.8$^{+2.3}_{-1.4}$& 1.06$^{+0.29}_{-0.15}$& 0.74$^{+0.36}_{-0.59}$\\  
D & 23:04:56.89 +12:19:52.6 & 1.02& 7.97 $\pm$ 1.93& 2.49 $\pm$ 0.03& 1.02&1.1 & 22.3$^{+2.6}_{-0.0}$& 1.77$^{+0.01}_{-0.72}$& 0.48$^{+0.46}_{-0.00}$\\ 
E & 23:04:55.82 +12:19:48.7 & 0.63& 3.64 $\pm$ 0.80& 0.44 $\pm$ 0.02& 0.61&1.2 & 21.2$^{+0.7}_{-1.0}$& 0.72$^{+0.09}_{-0.00}$& 0.89$^{+0.13}_{-0.30}$\\  
F & 23:04:56.75 +12:19:38.8 & 2.19& 6.77 $\pm$ 0.72& 4.72 $\pm$ 0.21& 1.72&1.3 & 22.5$^{+1.3}_{-2.1}$& 2.83$^{+0.21}_{-0.01}$& 0.61$^{+0.01}_{-0.07}$\\  
G & 23:04:56.65 +12:19:23.2 & 1.36& 9.66 $\pm$ 0.46&13.12 $\pm$ 0.35& 1.25&0.8 & 23.0$^{+2.5}_{-0.3}$& 1.83$^{+0.44}_{-0.45}$& 0.40$^{+0.06}_{-0.20}$\\ 
H & 23:04:56.61 +12:19:07.2 & 2.41&12.66 $\pm$ 0.74& 4.70 $\pm$ 0.18& 2.46&1.2 & 23.9$^{+0.7}_{-0.9}$& 3.48$^{+0.69}_{-0.29}$& 0.31$^{+0.01}_{-0.03}$\\ 
I & 23:04:57.34 +12:18:57.9 & 0.79& 6.99 $\pm$ 0.08& 0.51 $\pm$ 0.01& 0.74&1.1 & 22.1$^{+1.6}_{-0.4}$& 0.94$^{+0.01}_{-0.01}$& 0.52$^{+0.01}_{-0.01}$\\ 
K & 23:04:55.19 +12:18:57.7 & 0.48& 3.88 $\pm$ 0.38& 0.25 $\pm$ 0.01& 0.44&0.4 & 21.8$^{+2.0}_{-0.6}$& 0.85$^{+0.28}_{-0.20}$& 0.15$^{+0.10}_{-0.33}$\\  
L & 23:04:56.41 +12:18:54.2 & 1.36& 6.91 $\pm$ 0.67& 2.55 $\pm$ 0.08& 1.25&0.8 & 23.0$^{+2.5}_{-0.3}$& 1.83$^{+0.44}_{-0.45}$& 0.40$^{+0.06}_{-0.20}$\\  
M & 23:04:56.22 +12:18:35.6 & 0.92& 4.34 $\pm$ 0.26& 1.35 $\pm$ 0.03& 0.96&0.9 & 21.9$^{+1.5}_{-0.9}$& 0.89$^{+0.05}_{-0.22}$& 0.91$^{+0.23}_{-0.05}$\\
\enddata
%\vspace{-0.8cm}
\tablecomments{Each region consists of a circular aperture with a 15\farcs6 diameter and the reported center. All images were degraded to the angular resolution of the \cii observations. The FIR luminosity was computed by integrating the best-fitting model between 8 and 1000~$\mu$m. \cii and CO luminosities were computed by fitting the line inside the aperture, while the PAH luminosity was computed by subtracting the stellar component from the IRAC 8~$\mu$m photometry with the help of synthetic photometry from the best-fitting MagPhys model. Errors of FIR and PAH luminosities are less than 5\%. The metallicity Z is the value of the MagPhys model that best fits the photometric data.  Temperature of the ISM, stellar mass, and specific star formation rate are MagPhys outputs with 95\% confidence interval uncertainties.}
\end{deluxetable*}

\subsection{Herschel Observations}

NGC~7479 was observed by {\it Herschel} with the PACS and SPIRE instruments. In this paper we will consider only the PACS observations obtained at 70, 100, and 160~$\mu$m and the 250~$\mu$m image obtained with SPIRE. We do not consider the other SPIRE images, since their spatial resolution is too low in comparison to our SOFIA data, and they do not allow us to discern the emission from the different parts of the bar. Moreover, the emission at those wavelengths is well beyond the peak of the infrared emission and not useful to constrain the total infrared emission. We did not reprocess the images for our work since the quality of the archived products in the {\it Herschel} Science Archive is adequate for our analysis.

\subsection{H$\alpha$ Data}

We present data from H$\alpha$ Fabry--Perot observations, kindly provided by Stuart Vogel and Michael Regan. These observations were made with the Maryland--Caltech Fabry--Perot Spectrometer attached to the Cassegrain focus of the 1.5~m telescope at the Palomar  Observatory~\citep{Vogel1995}. The data  were  obtained on 1994 September 29--30. Forty exposures were taken, each with a 500~s integration time and a pixel scale of 1\farcs88. To improve the signal-to-noise  ratio, these data have been smoothed to 3\farcs6 resolution. The velocity planes are separated by 12.1~km~s$^{-1}$.

\subsection{UV, visible, Near-IR Data}

SEDs (spectral energy distributions) were created for parts of NGC~7479. We made use of observations in the two {\it GALEX} bands (FUV and NUV), SDSS images in the five Sloan bands ($u'$, $g'$, $r'$, $i'$, $z'$), and 2MASS images in the $J$, $H$, and $K_{\rm s}$ bands. The {\it GALEX} FUV observation are from the Nearby Galaxy Atlas survey. A deep NUV observation, taken in 2009 to observe the supernova SN 2009JF, was also used. Finally, we also analyzed a spectrum of the galaxy nucleus available in the SDSS archive. Images and spectra were retrieved from the respective archives: MAST, SDSS, and IRSA.

%\newpage
\begin{figure*}[!hbt]
\begin{center}
\includegraphics[width=0.98\textwidth]{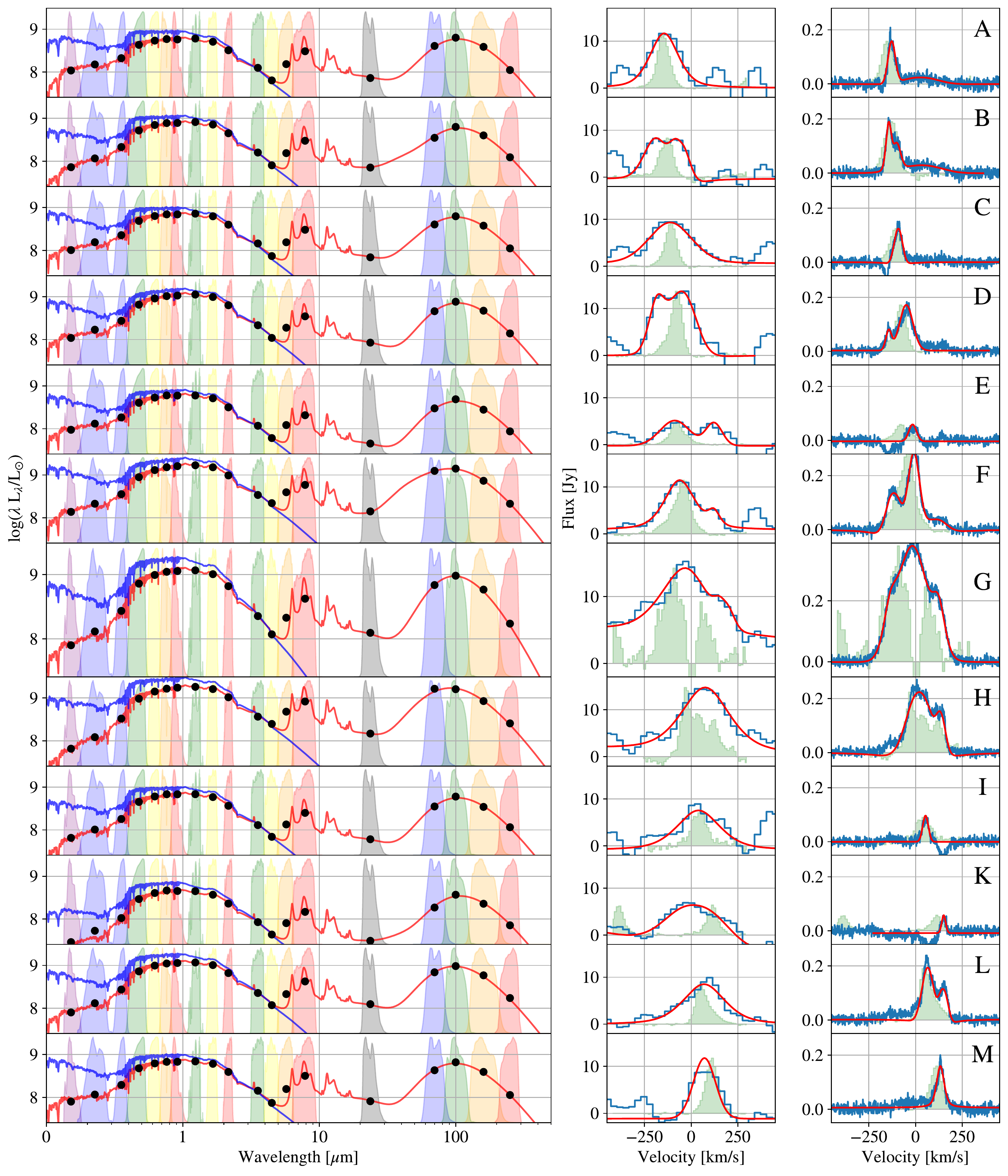}
\end{center}
\caption{SEDs, [\ion{C}{2}], and CO spectra for the 12 apertures considered. The spectral coverages of the various images used to estimate the SEDs (left panels) are shaded in different colors. The best fits obtained with MagPhys \citep{daCunha2008} are presented in red (attenuated distribution) and blue (unattenuated distribution). The [\ion{C}{2}] and CO spectra (middle and right panels, respectively) are plotted in blue. The red lines show the fits of continuum plus a combination of pseudo-Voigt functions. The green shaded spectra are the H$\alpha$ lines rescaled to the [\ion{C}{2}] and CO lines, respectively. The velocity shift along the bar is visible across the different apertures, from the top (North) to the bottom (South). 
}
\label{fig:sed}
\end{figure*}

\subsection{Chandra X-Ray Data}

The X-ray observations were retrieved from the Chandra archive. Two observations exist in the archive. They were obtained in August 11, 2009 to study the remnants of the supernova SN 1990U and, as a target of opportunity, two months later (November, 24) to follow up the more recent supernova SN 2009JF.  We reprocessed the data using the latest Chandra calibration ({\sc CALDB 4.9.3}) with {\sc CIAO}, version 4.12.1. To obtain an image we used the most recent observation (ID 11230) which has the longest exposure time (25~ks). For spectral extractions we considered also the other observation (ID 10120) which has an exposure time of 10~ks. The image, including photons between 0.3 and 8~keV, has been smoothed with an adaptive Gaussian kernel using the {\sc CIAO} tool {\sc dmimgadapt}. This routine smooths each pixel on optimal scales, in our case between 1 and 10 pixels. This is in order to reach the desired count threshold under the convolution kernel, in this case 10 counts. The resulting spatial resolution varies according to the counts, and it is better than 5~arcsec over the entire image. Because of the low photon statistics, we extracted spectra from the two observations and studied the combined spectrum after a separate background subtraction. We used the {\sc specextract} tool to extract X-ray spectra in the 0.3–8.0~keV range for the region centered on the nucleus and the regions on the counter-arms. The background was evaluated in a region close to the galaxy without X-ray emission in the same chip. The tool automatically scales the ratio to the same area when subtracting. We filtered the events based on the energy range and then grouped them to a minimum of 10 counts per bin prior to modeling the spectrum. Fluxes were estimated based on the count rates and the best fitting models.

\subsection{ALMA CO Data}
We used unpublished archival observations of the $^{12}CO_{J=1\rightarrow0}$ line in the 2.61--2.63~mm wavelength range taken with ALMA in band~3 (program ID: 2016.2.00195.S, P.I. Tanaka). The data were taken on 2017 September 21 and cover the whole region observed with SOFIA (see Fig.~\ref{fig:coverage}). NGC~7479 was observed for a total of 5136~s with the 7~m array for an expected line sensitivity of 17.6~mJy/beam. The spatial beam is an ellipse with axes of 15\farcs6 and 8\farcs0 and major axis oriented along the R.A. direction. The spectral resolution is 1.27~km/s, roughly ten times better than that of the data used by \citet{1999ApJ...511..709L}.

\begin{figure*}[!ht]
\begin{center}
\includegraphics[width=0.85\textwidth]{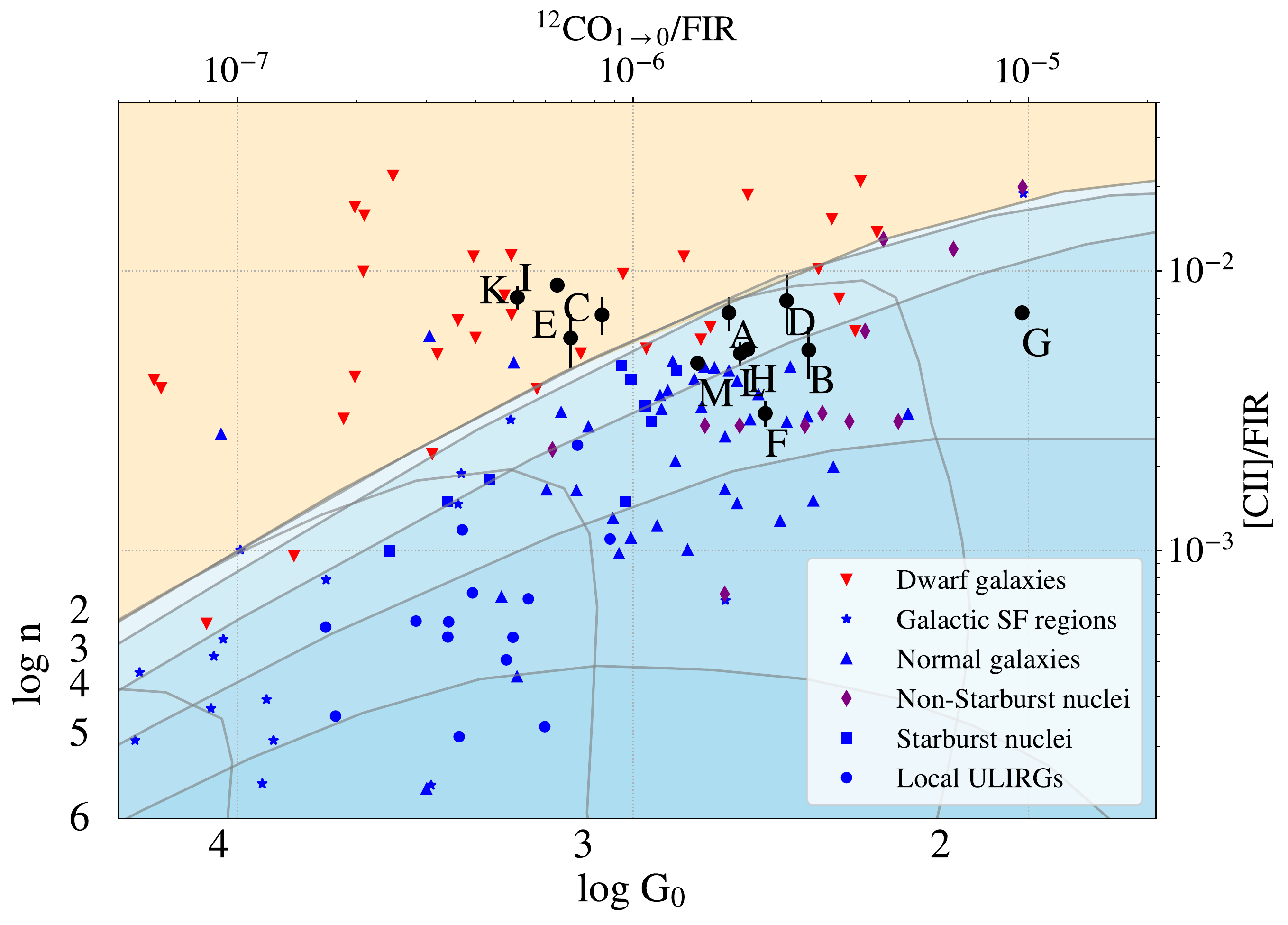}
\end{center}
\caption{
[\ion{C}{2}] vs CO relationship for local galaxies from~\cite{Madden2020} with a grid of PDR models as a function of gas density $n$ and the strength of the incident FUV field $G_0$ \citep{Kaufman1999}. Regions with \cii associated with PDRs are shaded in blue. The region shaded in yellow has either a \cii excess (e.g., due to excitation by shocks) or anomalously low CO emission (e.g., in sub-solar metallicity regions). The regions of NGC~7479 discussed in this paper are marked with black circles. The nucleus of NGC~7479 (region G) is less luminous in [\ion{C}{2}] than a typical star formation region. Regions I, E, K, and C emit more [\ion{C}{2}] than normal star formation regions and fall in the region populated by dwarf and low-metallicity galaxies.
}
\label{fig:ciico}
\end{figure*}

\section{Results and Discussion} \label{sec:results}

\subsection{Origin And Distribution of Gas Emission}
\label{subsec:origin}

In this section we study the distribution of the CO and \cii emission, and we try to relate the emission to the mechanism that produced it. To achieve this we compare the \cii, CO, and H$\alpha$ emissions at different locations in the galaxy to the broad-band emission from radio to X-ray wavelengths. In Fig.~\ref{fig:multiwave} we show the emission at different key wavelengths compared to the integrated emission from  \cii, CO, and H$\alpha$.

From \citet{Laine1998} we know that there is very little neutral atomic hydrogen along the bar of NGC~7479. Also, the H$\alpha$ emission is associated with star forming regions, as can be seen from the close spatial correspondence between the H$\alpha$ emission and UV emission intensities of the galaxy. On the other hand, a quick look at the integrated emission of CO and \cii presents a different picture of the galaxy, not directly correlated with the unobscured star formation.

To study the relationship between gas emission intensities at different wavelengths, we defined 12 different apertures with a diameter of 15.6~arcseconds, equivalent to the spatial resolution of our \cii map. Each aperture covers a different part of the bar and is centered either on a peak of the far-IR emission, or on a region with 20~cm radio continuum emission. At the top end of the bar we defined two apertures since the peak of the CO emission is displaced to the East with respect to the FIR emission (apertures B and C). Apertures A and M are situated at the locations where the bar meets the arms of the spiral galaxy. Aperture K is the only one defined outside of the bar in a region with H$\alpha$ and \cii emission, but with very low CO emission. The apertures are marked in Fig.~\ref{fig:multiwave}.

For each one of these apertures we measured the flux densities in the two GALEX bands, five SDSS bands, four IRAC bands, the MIPS 24~$\mu$m image, the PACS 70, 100, and 160~$\mu$m images and the SPIRE 250~$\mu$m image. To compare the emission at these bands with the \cii emission, we degraded the spatial resolution of each image to that of the \cii spectral cube. The spectral energy densities (SEDs) are presented in Figure~\ref{fig:sed}. Each SED has been fitted with the Magphys code~\citep{daCunha2008}. In the figure, the two lines represent the best fit with the attenuated and unattenuated distributions in red and blue respectively. In the same figure, the middle panel shows the \cii line from the apertures fitted with one or two pseudo-Voigt functions (red lines). The right panel, finally, shows the CO lines fitted again with a combination of pseudo-Voigt functions. The spatial resolution of the CO spectral cube has been degraded to the resolution of the \cii cube to have meaningful comparisons. The profile of the H$\alpha$ line, rescaled to the \cii and CO lines, is shown with a green shade.

The measurements of the intensity of the CO and \cii lines in the different apertures, as well as the main outputs from the code are reported in Table~\ref{tab:regions}.

\subsection{Relative Strength of CO and \cii Emissions}

When comparing the \cii and CO emissions, the main difference is the low luminosity in \cii at the nucleus of the galaxy (aperture G). This is especially evident when comparing the integrated emission shown in Figure~\ref{fig:multiwave}. The \cii emission is also extending towards the Southern end (aperture I) of the S-shaped radio continuum emission, unlike CO. Figure~\ref{fig:ciico} shows the relationship between the [\ion{C}{2}] and CO emission in the different apertures, normalized to the far-IR emission. This ratio can be used as a star formation diagnostic. 
% Ratio [\ion{C}{2}]/FIR vs Sigma(FIR)
\begin{figure}[!hbt]
\begin{center}
\includegraphics[width=0.48\textwidth]{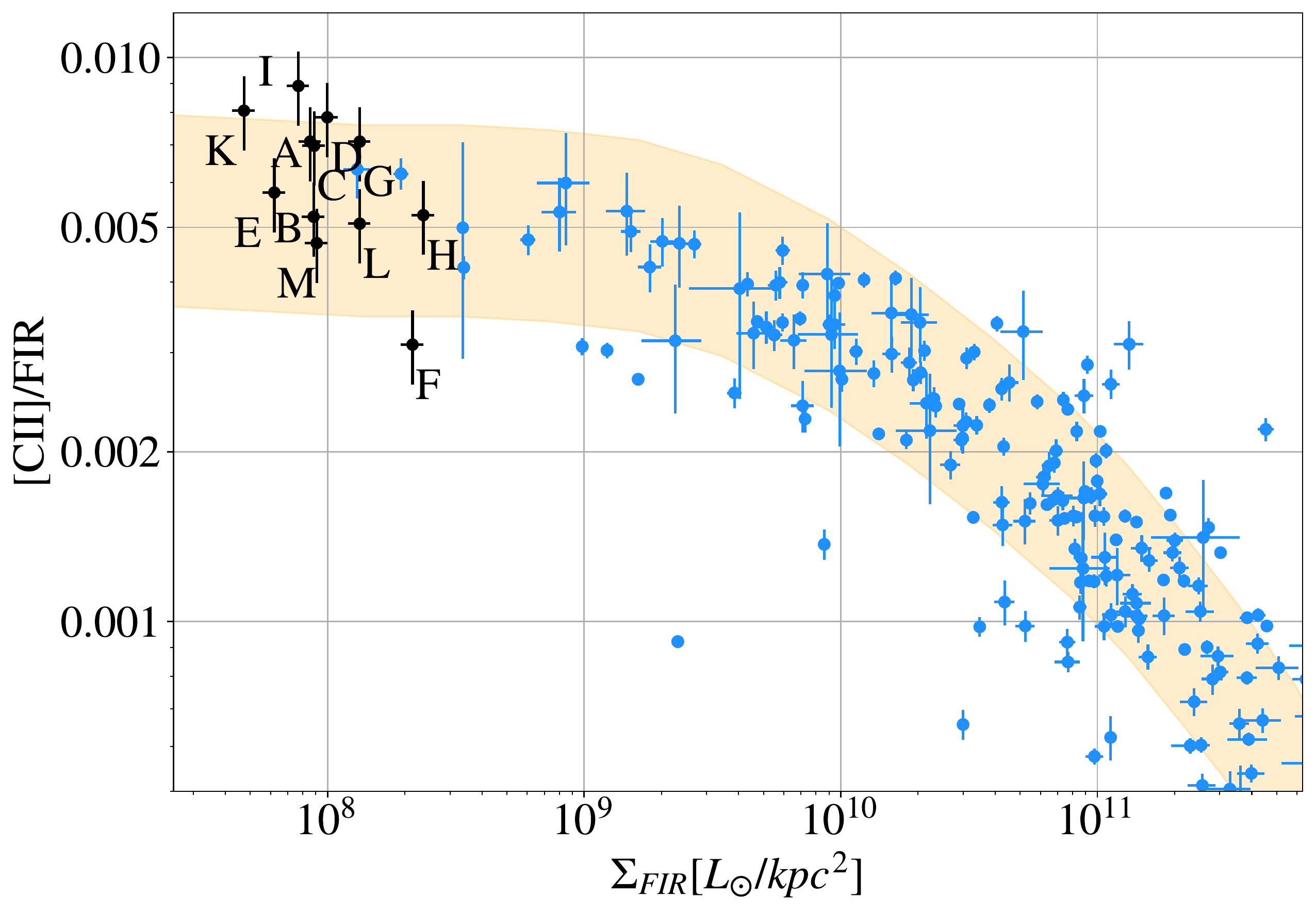}
\end{center}
\caption{
Ratios of \cii over far-IR emission as a function of the infrared surface brightness for the 12 apertures considered. The blue symbols correspond to the comparison sample from the GOALS project \citep{DiazSantos2017}. The shaded region is the fitted curve with 1-$\sigma$ uncertainty. Region I, corresponding to the radio/X-ray hot spot at the southern end of the jet, shows an excess of \cii emission.
}
\label{fig:ciifir}
\end{figure}
We show, for reference, a grid of PDR models from \citet{Kaufman1999}. The region shaded in blue corresponds to emission possible with pure PDR models. The region shaded in yellow cannot be explained in terms of pure PDR emission, but requires either an excess of \cii emission or a deficit of CO emission. Most of the normal and star forming local galaxies lie in the blue region, while higher ratios are measured for dwarf galaxies and lower metallicity galaxies~\citep{Madden2020}.

The regions at the ends of the radio continuum jet (apertures I and E) have a ratio higher than normal star forming regions. The same is true about region C that we already pointed out as having very little CO emission and region K which lies outside of the bar.

 Regions I and E are especially interesting because they seem to have only very faint and narrow H$\alpha$ and CO emission, whereas the \cii emission is double-peaked in the case of Region E, and broad in the case of Region I. These regions correspond to the ends of the radio counter-arms. In this case there are strong similarities with NGC~4258 \citep{Appleton2018}, where evidence was presented that the enhanced \cii emission traces the dissipation of mechanical energy through shocks and turbulence as the jet interacts with the surrounding ISM. \citet{Lesaffre2013} showed that even quite low-velocity shocks, passing through a mildly UV-irradiated diffuse (10$^2$--10$^3$ cm$^{-3}$) molecular medium, can produce strong \cii emission, comparable to other powerful ISM coolants, like mid-IR H$_2$ emission. Models of this sort were used to explain the powerful H$_2$, \cii and H$_2$O emission detected by {\it Spitzer} and {\it Herschel} in the shocked filament in Stephan's Quintet \citep{Guillard2009, Appleton2013, Appleton2017}. A similar mechanism was put forward to explain the clear association of \cii emission with warm H$_2$ and faint soft X-ray emission associated with the end of the southern radio jet and anomalous radio arms in NGC~4258 \citep{Appleton2018}. The necessary mild UV radiation field required to ionize the carbon is provided by the general galactic stellar background.  In NGC~7479, although we do not have direct evidence of shock-heated warm molecular gas, the association with the X-ray emission (see Fig.~\ref{fig:xray}), and the unusual \Cii/FIR ratios discussed in the next subsection, are consistent with this picture.
%
% Ratio [\ion{C}{2}]/PAH(7.7um) vs  vFv(70)/vFv(100)
\begin{figure}[!t]
\begin{center}
\includegraphics[width=0.48\textwidth]{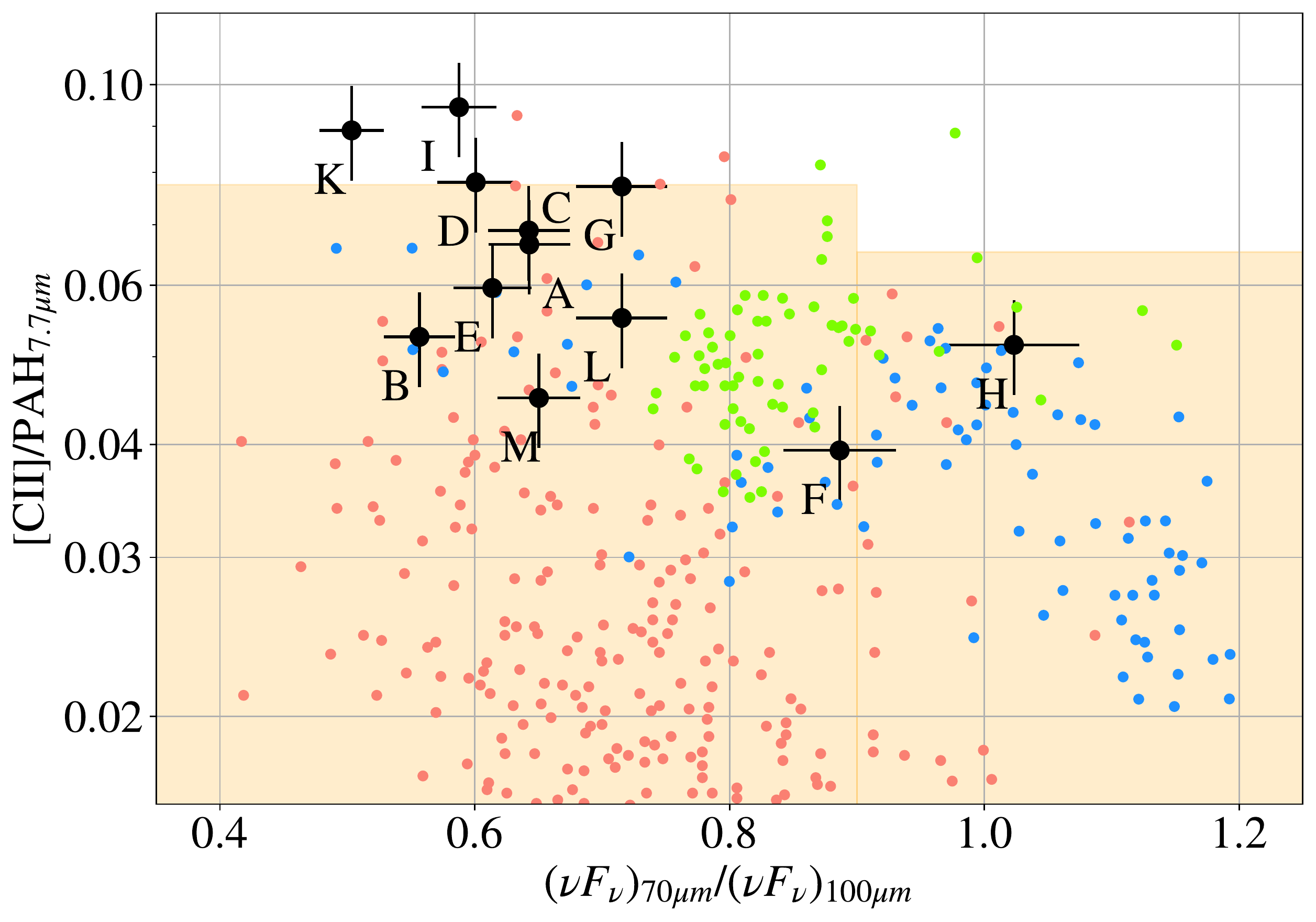}
\end{center}
\caption{
Ratio of \cii over PAH~7.7~$\mu$m emission as a function of the far-IR slope for the 12 apertures considered. The comparison sample consists of regions in NGC~1097 (blue) and NGC~4559 (green) from \citet{Croxall2012}, and in NGC~6946 (red) from \citet{Bigiel2020}. The shaded part of the plot contains 99\% of the regions in the three comparison galaxies. The highest ratio in NGC~7479 is found in region I that corresponds to the radio hot spot at the southern end of the radio continuum jet.
}
\label{fig:ciipah}
\end{figure}

\subsection{Infrared Diagnostics}

The relationship between \cii and star formation can be tested using mid- and far-IR diagnostics. The far-IR emission (between 8 and 1000~$\mu$m) has been shown to be a good estimator of star formation~\citep{Kennicutt1998}. Infrared surveys, such as the GOALS survey~\citep{DiazSantos2017}, found a good correlation between the \cii emission strength and the total far-IR luminosity. The ratio is fairly constant for normal galaxies, while ultra-luminous infrared galaxies have a deficit of \cii emission. The same relationship can be used to explore various regions of a galaxy to see if the \cii emission is related to star formation, or if there is an excess or deficit of such emission with respect to the star formation rate measured. For each aperture, we computed the FIR emission by integrating the spectral energy distribution of the best fitting MagPhys model~(see Fig.~\ref{fig:sed}). In Figure~\ref{fig:ciifir} we compare the values of the \Cii/FIR ratios in the different apertures on NGC~7479 with values from the GOALS sample. Most of the apertures lie in the region of normal galaxies, except for the aperture I, which lies on the southern end of the radio continuum jet emission.  The \Cii/FIR ratio is anomalously high at this location.

Another quantity that correlates very well with the star formation rate is the emission at 7.7~$\mu$m from the polycyclic aromatic hydrocarbons \citep[PAHs,][]{Peters2004}. Most of the \cii emission in normal galaxies originates from cooling of neutral gas in photo-dissociation regions~\citep{Croxall2017}. On the other hand, the main mechanism to heat the neutral gas is via photoelectric heating by interstellar PAHs \citep{Draine1978,Tielens1985,Hollenbach1989} which explains the good correlation between PAH and \cii emissions. Therefore, this ratio is very sensitive to the \cii emission mechanism. The 7.7~$\mu$m PAH luminosity is estimated by subtracting the stellar emission from the IRAC 8~$\mu$m image. The stellar emission estimate was computed through synthetic photometry of the unattenuated flux from the best-fitting MagPhys model. 

By comparing the values in different regions of normal star-forming galaxies \citep{Croxall2012,Bigiel2020}, we see that region I and K have anomalously high ratios. 
Region K that lies outside of the bar seems to have some \cii excess. It is possible that this region may be an example of "CO-dark" molecular hydrogen. This idea is supported by the observation that this region has the lowest best-fitting SED metallicity of all the regions observed (see Table~\ref{tab:regions}), a condition that may be conducive to "CO-dark" molecular gas \citep{Wolfire2010}. Recent studies of low metallicity regions in galaxies~\citep{Madden2020} show that most of CO is dissociated in these environments, making \cii a better tracer of molecular gas in these particular regions.
Region I, which lies close to the end of the continuum radio jet in the South, will be discussed in the next subsection in the context of possible ISM heating by the jet. We note that the corresponding region E (at the tip of the northern jet) 
does not seem to have anomalous ratios in these two diagnostics. However, we caution that there are two velocity components to the [CII] emission from Region E, and so any excess in this ratio in one component may be diluted by normal PDR emission from the other.

\subsection{The Counter-arm Structure}
\label{sec:jet}

\begin{figure}[!t]
\begin{center}
\vbox{
\includegraphics[width=0.48\textwidth]{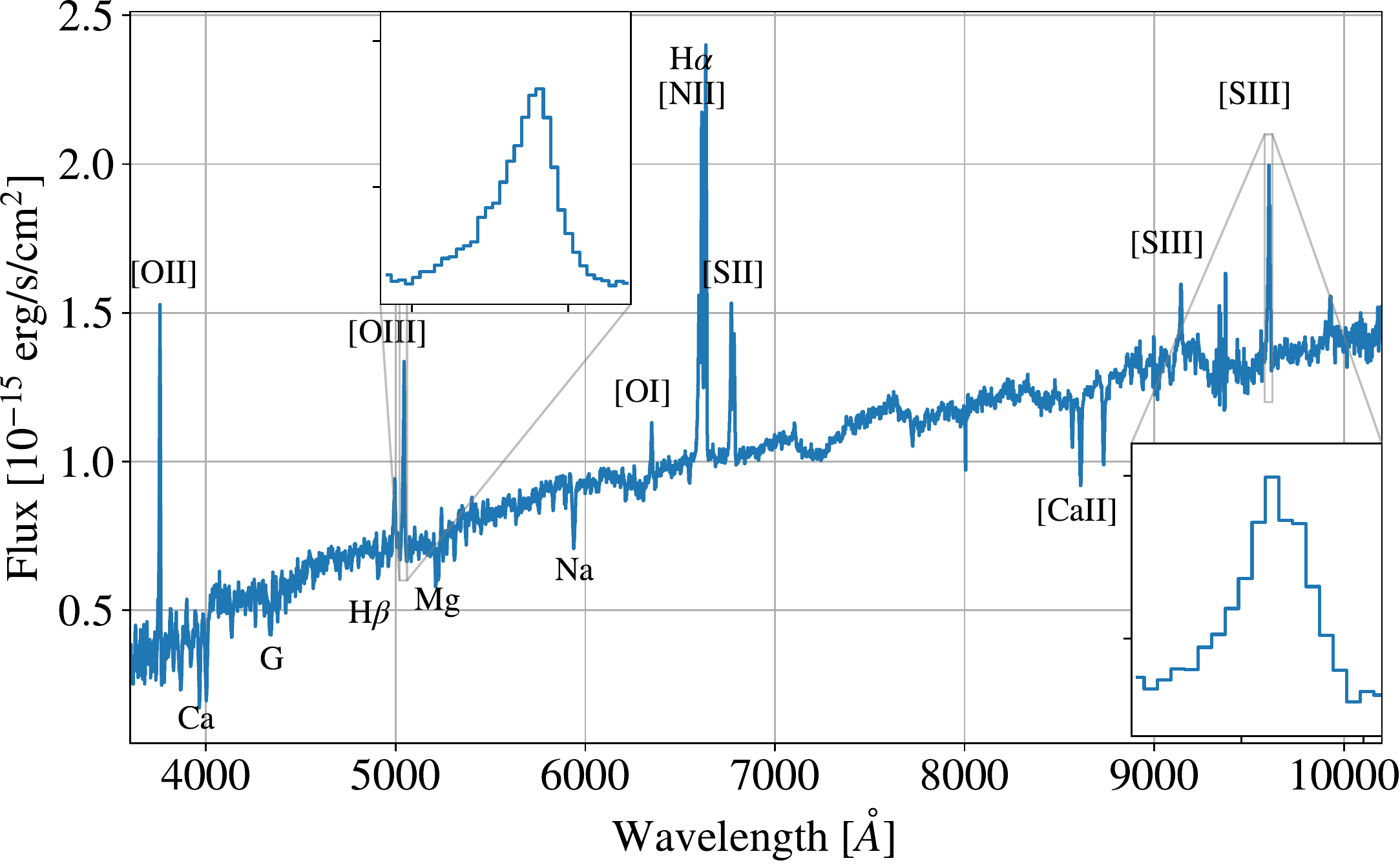}\\
\includegraphics[width=0.48\textwidth]{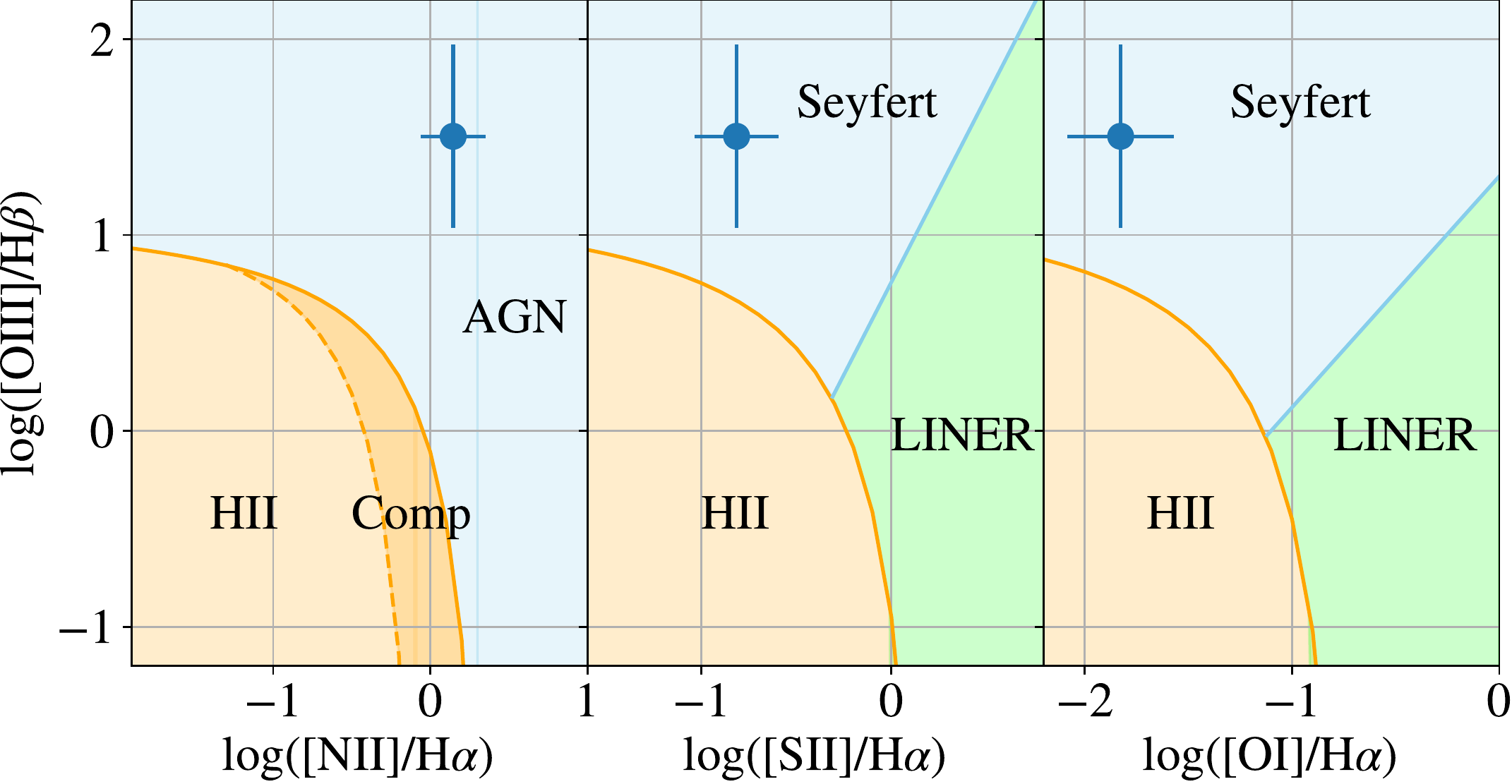}
}
\end{center}
\caption{Top: SDSS spectrum of the nucleus of NGC~7479 with main spectral features identified. The asymmetric profiles of the [OIII]5008\AA~and [SIII]9531\AA~lines are shown in the insets. Bottom: \citet{Kewley2006} diagnostic diagrams for the nucleus of NGC~7479. The galaxy, blue dot with errorbars in the figures, can be safely classified as a Seyfert.
}
\label{fig:sdssspec}
\end{figure}

NGC~7479 is known to host an active nucleus. \citet{Ho1997} classified its nucleus as a Seyfert~1.9, although previously it was classified as a LINER~\citep{Keel1983}. A recent SDSS spectrum of the nucleus (observed in March 2012, see Fig.~\ref{fig:sdssspec}) shows clearly that the galaxy can be classified as a Seyfert according to the BPT diagrams by \citet{Kewley2006}. From the line ratio H$\beta$/[OIII]5008\AA~$=0.22\pm0.1$ the galaxy can be classified as Sy 1.8, according to the schema of \citet{Winkler1992}.
The SDSS spectrum shows a high extinction ($A_V = 8.4$~mag, from the Balmer ratio decrement) typical of Seyfert type 1.8--2 galaxies. Moreover, blue wings are visible in narrow lines such as [OIII]5008\AA~and [SIII]9068\AA~(see insets in the top panel of  Fig.~\ref{fig:sdssspec}). Such asymmetric line profiles are usually associated with outflows of gas \citep[see, i.e., ][]{Schmidt2018}.

\begin{figure*}[!t]
\begin{center}
\hbox{
\includegraphics[width=0.44\textwidth]{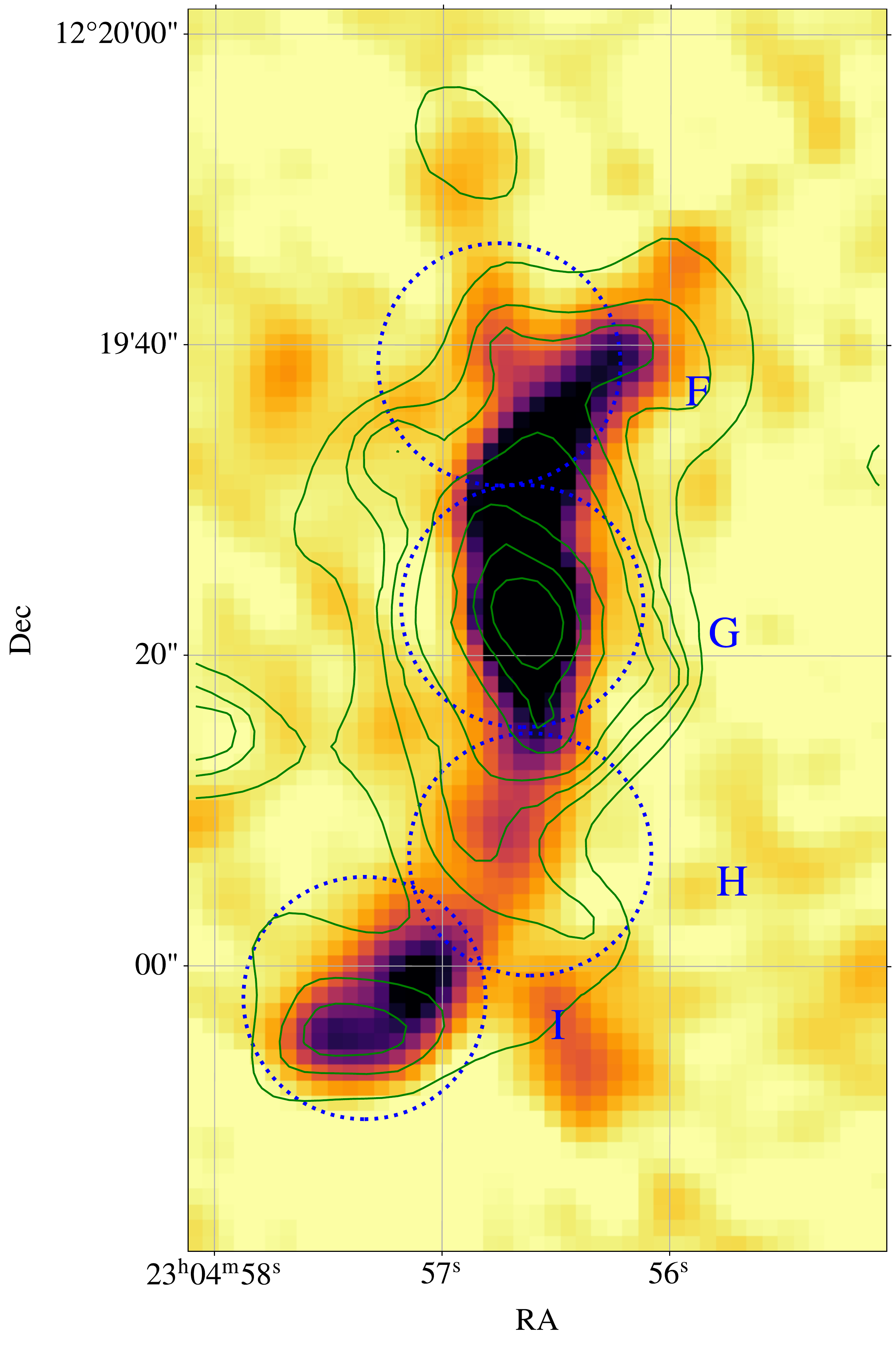}
\hspace*{-4.5cm}
\vbox{
\includegraphics[width=0.36\textwidth]{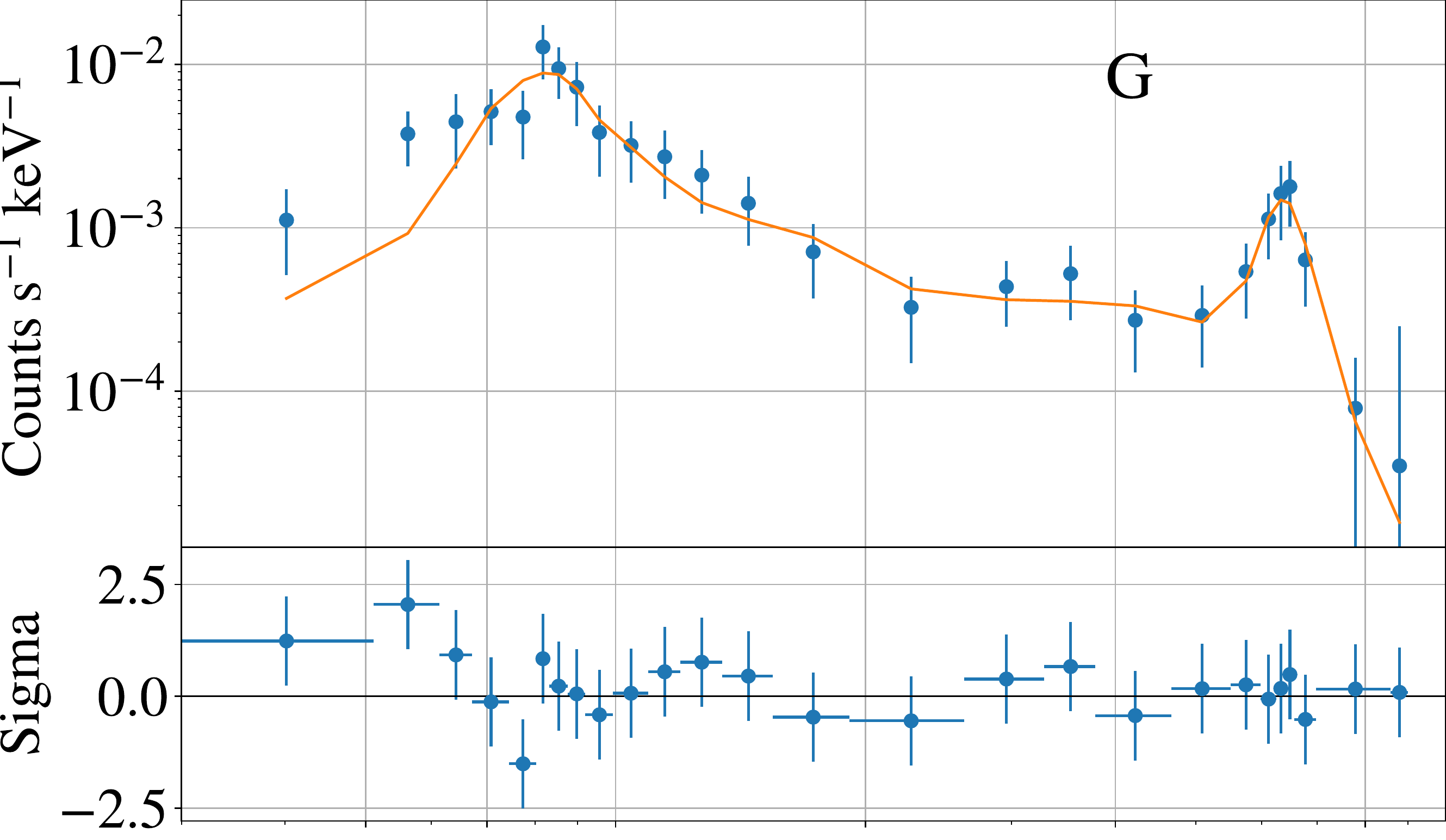}\\
\includegraphics[width=0.36\textwidth]{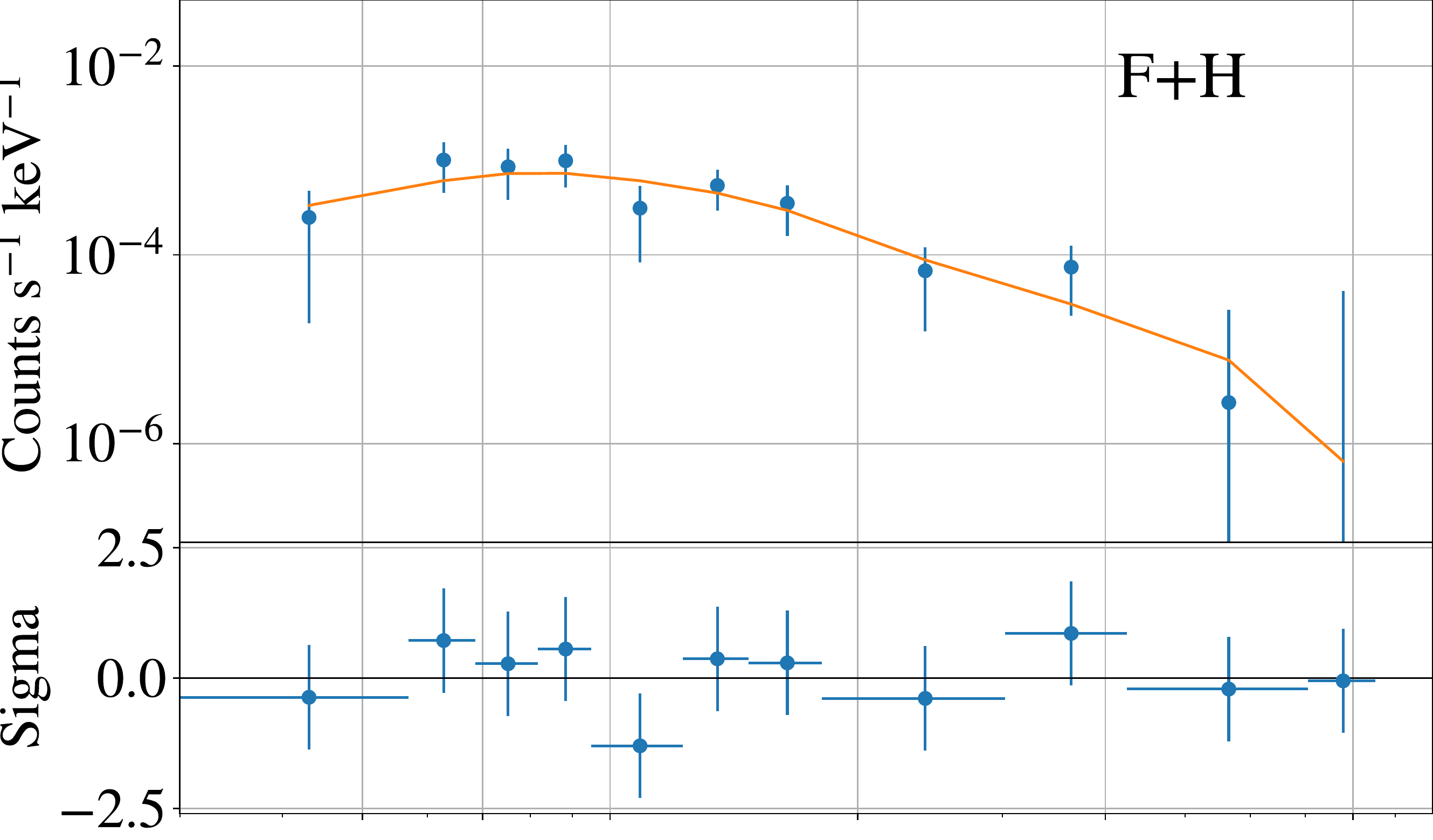}\\
\includegraphics[width=0.36\textwidth]{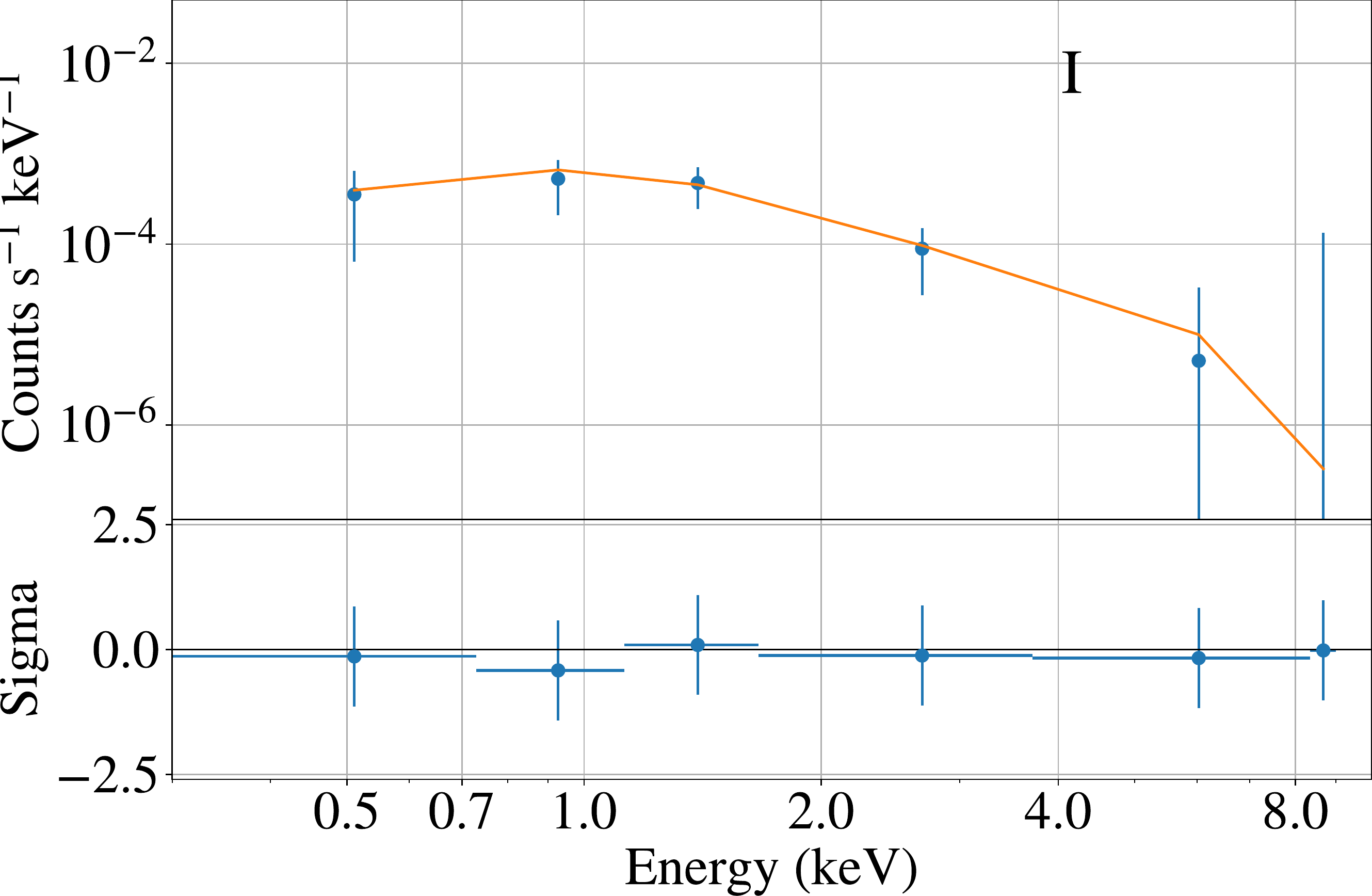}
}
}
\end{center}
\caption{
On the left, contours of the 0.5--8~keV X-ray emission over the 20~cm radio continuum map \citep{Laine2008}. The contours follow a pattern similar to that of the radio intensity, confirming the nuclear origin of the radio emission. In particular, the southern end of the jet-like structure is clearly detected in the X-ray image, although it peaks at a slightly different location. On the right, spectra extracted in the apertures traced with dotted circles in the image. The models fitting the data are traced with orange lines, while the residuals of the fits are shown at the bottom of each spectrum.
}
\label{fig:xray}
\end{figure*}
Radio observations at 20~cm by \citet{Laine2008} showed evidence of the existence of a jet-like structure with arms opening in a direction opposite to the optical arms. They were not able to see this structure at any other wavelength and they speculated that this radio emission could be linked to a jet emanating from the nucleus of the galaxy.

Our reprocessing of Chandra archival data shows that the X-ray emission follows the same pattern as the counter-arms visible in the 20~cm radio continuum. As shown in Fig.~\ref{fig:xray}, the X-ray contours follow the same orientation as the radio structure. Moreover, the southern end of the structure  that has the strongest radio emission, is also a hot spot in the X-ray observations. NGC~7479 is considered to harbor an AGN on the basis of optical spectra of its nucleus. In Fig.~\ref{fig:xray} we show the X-ray spectrum extracted from aperture~G that shows a prominent Fe K$\alpha$ feature at 6.4~keV. We used the {\sc sherpa} package of {\sc CIAO} to model the spectrum. We obtained a good fit with a combination of a power law and {\sc apec} thermal models from the {\sc sherpa} library considering the intrinsic absorption as a free parameter. The Fe K$\alpha$ line at 6.4~keV was fitted with a Gaussian function. The spectrum is typical of an AGN: it has a high hardness ratio\footnote{The hardness ratio is defined as $HR = \frac{H-S}{H+S}$ with $H$ and $S$ the hard (0.5--2~keV) and soft (2--8~keV) fluxes, respectively.} ($HR =  0.8 \pm 0.5$), and the Fe K$\alpha$ line at 6.4~keV is clearly detected. Moreover, the high \ion{H}{1} absorption required for a good fit (N$_H = 0.8^{+0.1}_{-0.3}\times10^{22}$~cm$^{-2}$) and the large width of the 6.4~keV line (FWHM$ = 0.9^{+0.2}_{-0.7}$~keV) suggest that the galaxy harbors at its center an heavily obscured active nucleus. This analysis confirms previous results obtained with 13~ks XMM observations \citep{Wang2010}. The ratio of our estimated values of optical extinction ($A_V$) and \ion{H}{1} absorption (N$_H$) are not far from the Galactic standard ratio \citep[see Fig.3 in][]{Burtscher2016}. The rest of the X-ray emission is much weaker as shown in the two other spectra in Fig.~\ref{fig:xray}. Also, the hardness ratios of the X-ray emission in the other apertures are much lower than in the nucleus ($HR = 0.0$ and $-0.2$ for apertures I and F+H, respectively). We fitted the other two spectra either with {\sc apec} thermal models or a combination of power-law and {\sc apec} thermal models. In both cases we also assumed a Galactic intrinsic absorption, obtaining similar results. With these models we can estimate the 0.5--8~keV flux inside the southern X-ray hot spot (aperture I) and the average flux inside the two other apertures along the bar (F and H). The flux inside aperture I is $5_{-2}^{+2}\times 10^{-15}$~erg s$^{-1}$cm$^{-2}$, while the average flux inside apertures F and H is $5_{-1}^{+2}\times 10^{-15}$~erg s$^{-1}$cm$^{-2}$. By considering the relationship between X-ray emission and the total IR flux in \citet[][their equation 23]{Mineo2012} for normal star-forming galaxies, we find that aperture I has a ratio of 1.4 between the expected IR emission (based on the X-ray emission) and the measured IR emission. For our apertures that lie on the bar (F and H), the ratio is 0.5, i.e., approximately one third of the value found in aperture~I. If we take into account the fact that most of the IR emission in aperture~I is located close to the bar while the X-ray emission peaks on the opposite side, we conclude that at least some of the X-ray emission cannot be explained with star formation only. It is therefore reasonable to assume that the S-like structure detected in the radio and X-ray is associated with emission from the AGN. Similarly to another galaxy showing radio counter-arms \citep[NGC~4258, see][]{Appleton2018}, this structure can be explained by invoking the existence of a jet originating inside the nucleus, and colliding with dense clumps of gas along the bar~\citep{Plante1991,Daigle2001,Mukherjee2016}. If the jet is emitted at an angle with respect to the bar, during the collisions the gas transfers momentum to the clouds of gas along the direction of the bar, hence gradually changing the jet direction as the component of velocity along the bar decreases. As the jet exits the bar and enters the less dense disk region, the direction of the jet remains constant. This scenario can explain the shape of the counter-arms.

\label{subsec:radiojet}
\begin{figure}
\begin{center}
\includegraphics[width=0.42\textwidth]{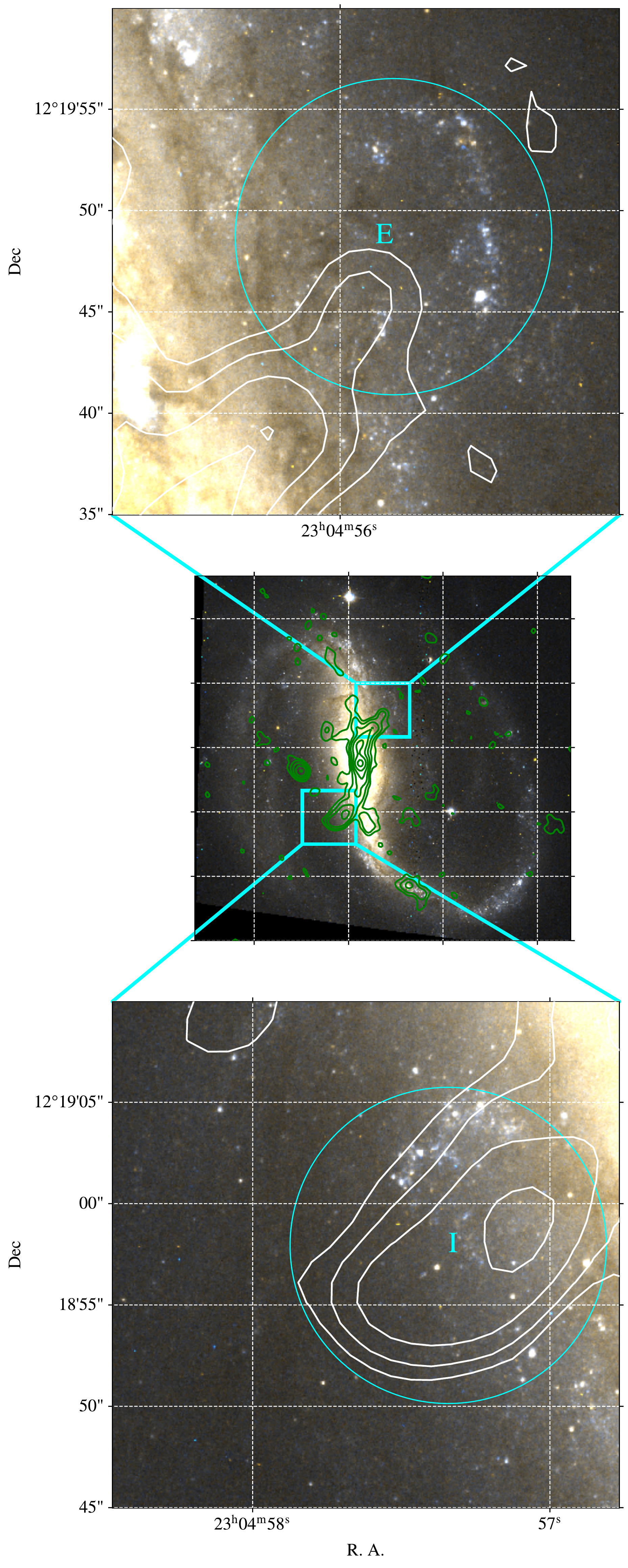}
\end{center}
\caption{
End locations of the radio jet-like emission in NGC~7479. The middle panel shows the HST two-color image with 20~cm radio continuum logarithmic contours from 0.1 to 4~mJy/beam. The top and bottom panels show the regions at the end of the radio jet-like emission. The cyan circles are the apertures E and I considered in this paper. In the top panel, the tip of the radio emission is surrounded by bright young stars. The bottom panel shows that the southern end of the radio jet has new stars on the top and below the radio emission peak. The jet likely exits the disk since no star formation is visible at its end.
}
\label{fig:radiojet}
\end{figure}

Figure~\ref{fig:radiojet} shows the comparison between visible and radio emission at the ends of the jet-like structure and the apertures used to measure the CO and \cii emission in this paper. As shown in the previous sections, the ratio of the \cii to CO emission is anomalously high in these two regions with respect to regions of \begin{figure*}[!hbt]
\begin{center}
\includegraphics[width=\textwidth]{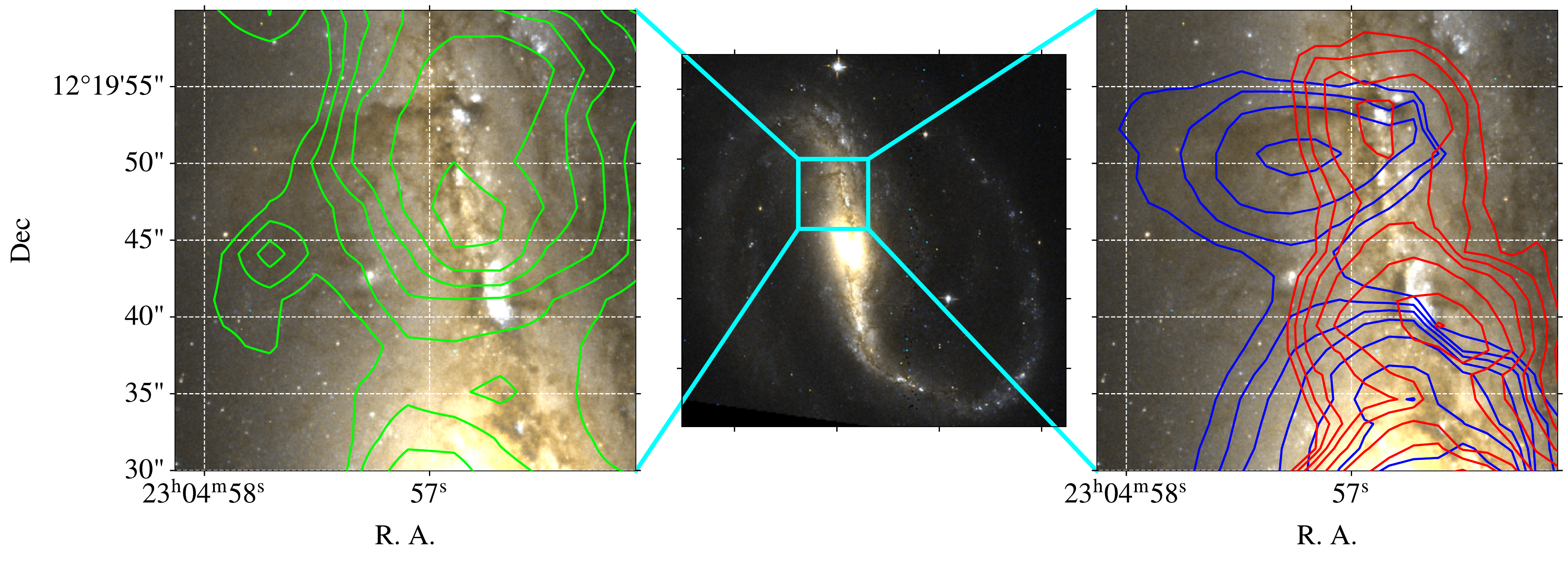}
\end{center}
\caption{Region with remnants of a possible minor merger (nucleus of a merging galaxy, tail of star formation, horizontal dust lane) in the two-color {\it HST} image. The contours of the total [CII] emission are shown in green in the left panel. The two velocity components of the CO emission are shown in the right panel. While the higher velocity component (red, $\Delta v \approx -50$~km/s) follows the merging structure with the maximum on the nucleus, the lower velocity (blue, $\Delta v \approx -150$~km/s) component shows an extension aligned with the horizontal dust lane crossing the bar. The [CII] emission also shows two velocity peaks, but the spectral resolution is not sufficient to distinguish between the two components across the entire image.
}
\label{fig:merging}
\end{figure*}
normal star formation. However, when using mid-IR and far-IR diagnostics, the northern region (E) seems compatible with star formation, while the southern region (I) shows an excess of \cii emission.

To better understand the reason for this behavior, it is instructive to look at the clusters of young stars located at the ends of the radio continuum arms. At the northern end, the radio emission points to a region which is opaque in visual wavelengths, and is surrounded by a region full of clusters of bright blue stars. The same region (region E) is also bright in the ultraviolet (see Fig.~\ref{fig:multiwave}).  This morphology is consistent with at least the majority of the  \cii emission originates in star forming regions. This might explain why the \cii emission correlates very well with mid-IR and far-IR estimators of star formation. On the other hand in this region, we cannot rule out some fraction of the [CII] emission originating in warm molecular gas heated by the jet, since the [CII] line profile is double-peaked, but the CO emission is not.

At the southern end, on the contrary, the jet is only partially surrounded by star clusters. The northwestern edge of the southern jet end has a front of bright young stars, and just in front of the maximum radio emission there is an arch-like cluster of blue stars. Beyond these stars, there seems to be little star formation associated with the jet. This is exactly where the peak of the X-ray emission is located. The impression in this case is that the jet is coming out of the disk, and therefore, has no possibility to interact with the dense gas in the disk. Going out of the disk, the jet is still able to interact with lower density molecular gas in the halo, which probably triggers a more intense X-ray radiation. Even if the halo gas density is not high enough to trigger star formation, the energy of the jet dissipates by shocking the molecular gas that later cools down, emitting [\ion{C}{2}]. This scenario would explain the excess [\ion{C}{2}] emission with respect to the lower PAH and FIR emission. The lack of H$\alpha$ and \ion{H}{1} emission in the region supports this hypothesis.

\subsection{Merging Remnants}
\label{sec:merging}

\citet{Laine1999} were able to explain the shape and features of NGC~7479 with a minor merger model. In such a model, a region north of the nucleus contains visible remnants of the merging process. In particular, what is left of the nucleus of the less massive galaxy captured by NGC~7479 is still visible just north of the nucleus. As shown in the {\it HST} image in Fig.~\ref{fig:merging}, the bright elongated nucleus is followed by a trail of forming stars, likely a residual of an arm. Moreover, a thick dust lane across the bar could also be a residual of the merged galaxy. The merging left the northern part of the galaxy in a much more turbulent state than its southern part. In Fig.~\ref{fig:merging} the contours of the \cii and CO emission are overlaid on an {\it HST} image. The region trailing the merging nucleus appears very bright in \Cii. As discussed in the next section, the velocity structure of \cii shows a double peak over this region. 
\begin{figure*}[!t]
\begin{center}
\includegraphics[width=\textwidth]{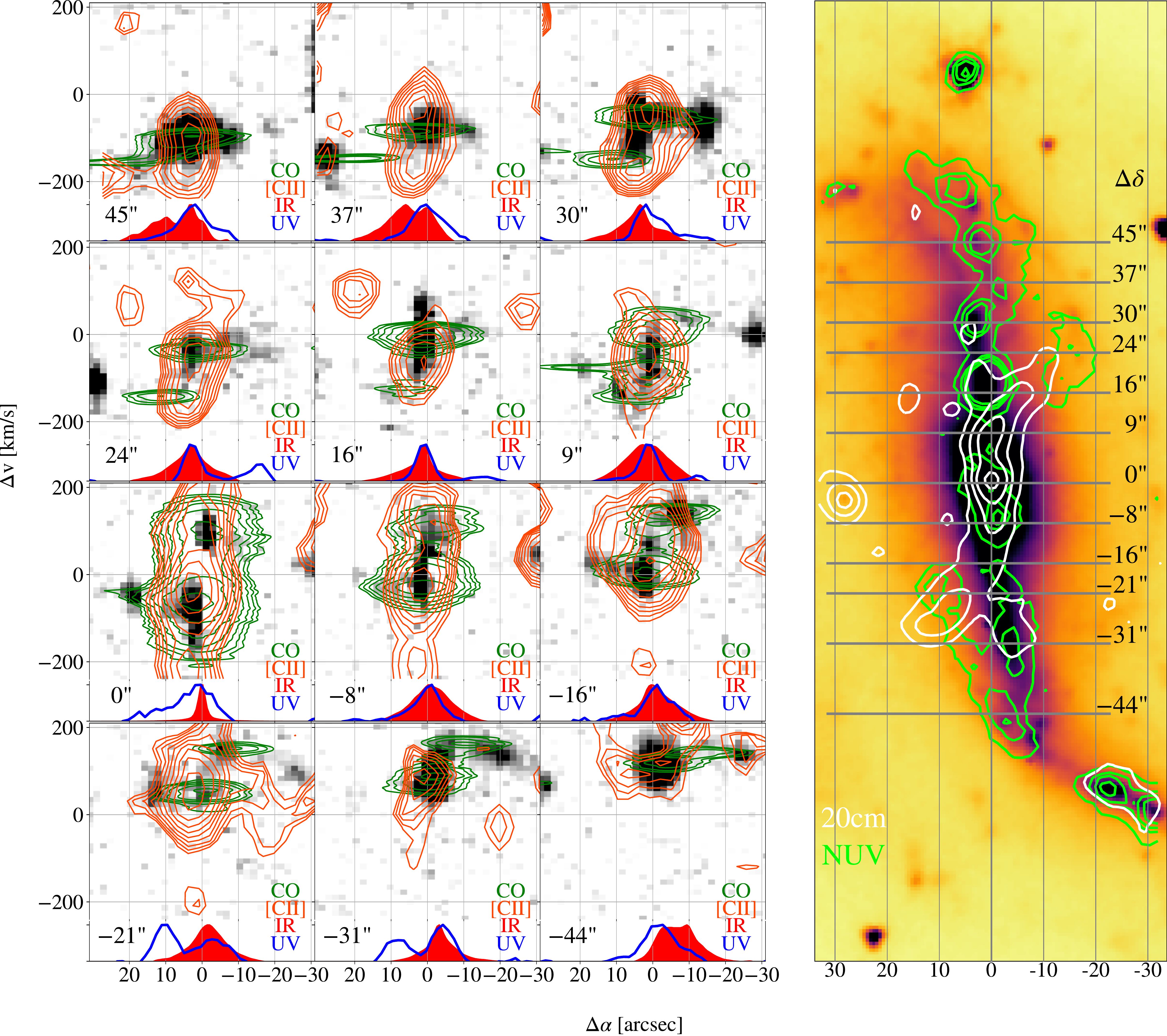}
\end{center}
\caption{On the left, slices in declination of the  H$\alpha$ emission (grey shaded) with overlapped logarithmic contours of the CO (green) and [\ion{C}{2}] (orange) emissions showing the distribution of the atomic and molecular gas in velocity and R.A. across the bar.  The image on the right shows the intensity of the 3.6~$\mu$m radiation with overlapping contours of the near-UV (green) and 20~cm radio continuum (white) emissions and a vertical grid of angular distance in arcseconds from the central position. Each slice corresponds to an horizontal segment on the image. The distance in declination of each slice from the nucleus is marked on the right side of each segment on the image and on the bottom left corner of each subplot. The normalized intensity profile of the UV and IR emission along the slices is shown at the bottom of each subplot in blue and red, respectively.
}
\label{fig:3wavtomography}
\end{figure*}
Unfortunately, the spectral resolution of FIFI-LS is not good enough to clearly separate the two components.

\begin{figure*}[!t]
\begin{center}
\includegraphics[width=0.75\textwidth]{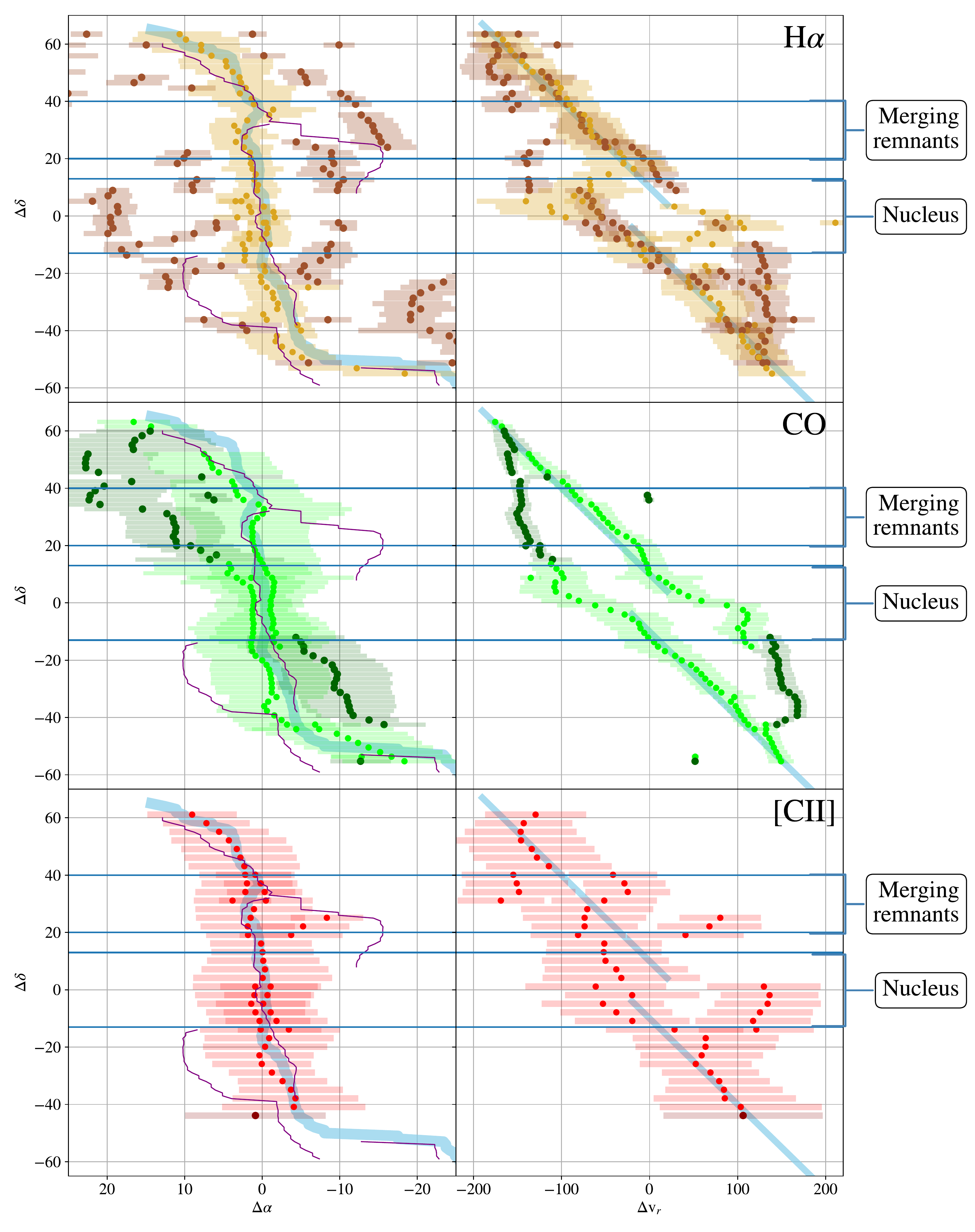}
\end{center}
\caption{Displacement in R.A. with respect to the center (left)  and line-of-sight velocity relative to the systemic velocity (right) for the H$\alpha$, CO, and [\ion{C}{2}] components.
The assumed values are: 23:04:56.63 +12:19:22.7 (J2000) for the galaxy center and 2381~km/s for the systemic velocity. In the left panels, the peak of the far-IR emission at 70~$\mu$m is traced with a wide light blue line and the locations of UV emission with a thin purple line. In the right panels, the projected rotational velocity of 5.6~km~s$^{-1}$~arcsec$^{-1}$ of the two sides of the bar is traced with a broad blue line to put in evidence the rigid rotation of the bar. The dots mark the peak emission of the components, while the horizontal bar shows their extents. Peaks more than 4~arcsec from the far-IR peak are marked with darker colors. The declination range of the nucleus and of the merging remnants are indicated on the right.
}
\label{fig:velra}
\end{figure*}

The CO emission also presents two clear peaks in velocity (see next section). In this case, thanks to the high spectral resolution of ALMA, it is possible to obtain the integrated intensities of the two components by fitting two velocity components over the entire spectral cube. In the region with merging remnants, the component with the velocity of $-50$~km~s$^{-1}$ with respect to the systemic velocity follows the dust lane of the bar. The emission is all over the merging structure and peaks on the merging nucleus. The other component, approximately at $-150$~km/s from the systemic velocity, traces a cloud of molecular gas which is limited by the horizontal dust lane. It is possible that molecular gas flows towards the bar following this dust lane. We conclude that the turbulent velocity profile of the \cii emission and the presence of a cloud of molecular gas along the dust lane are other possible tracers of an ongoing minor merger event.

%Tomography
\begin{figure}[!th]
\begin{center}
%\hbox{
\includegraphics[width=0.48\textwidth]{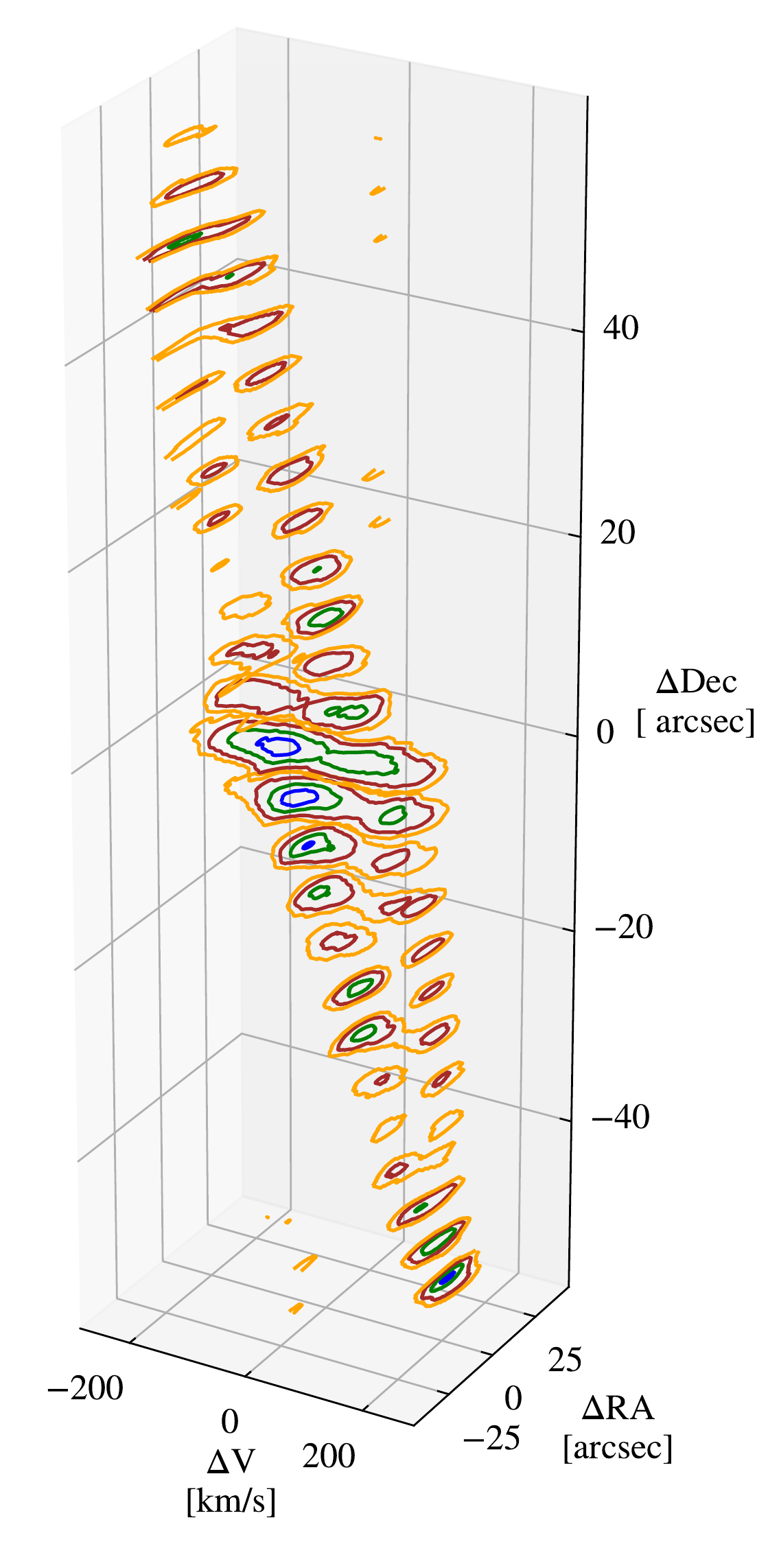}
%\includegraphics[height=22cm]{CO3d.pdf}
%\includegraphics[height=22cm]{COtomography.pdf}
%}
\end{center}
\caption{3D view of the molecular gas emission along the galaxy bar detected with the CO line. Isocontours of sections in declination of the CO spectral cube at 20, 30, 60, 80 $\mu$Jy/pixel are shown in yellow, red, green, and blue, respectively.
%On the left, the 3D representation of integrated cuts in declination with isocontours (0.03, 0.06, 0.1 mJy/pixel in yellow, red, and green, respectively). On the right, images of sections of the spectral cube at different declinations.
}
\label{fig:tomography}
\end{figure}

\subsection{Gas Kinematics}
\label{sec:kinematics}

A way to visualize the kinematics of the gas in NGC~7479's bar region is to consider sections in declination of the bar, since the bar is almost aligned along the north--south direction. By plotting the intensity of the spectral cube in the velocity--R.A. space, it is possible to identify different components of the emission. In Fig.~\ref{fig:3wavtomography} we display 12 different sections of the spectral cube of H$\alpha$, CO, and \cii emissions corresponding to interesting regions along the bar. The declinations at which we sliced the spectral cubes are displayed in the right column panel of the figure as horizontal segments over a near-IR image (the IRAC map at 3.6~$\mu$m) with green contours of the near-UV emission (GALEX image) and white contours of the 20~cm continuum (VLA). The image summarizes the main components of the bar: old stars (3.6~$\mu$m), new stars (near-UV), and radio counter-arms. For each declination, the normalized intensity of the IR and UV images as a function of the R.A. is displayed at the bottom of the corresponding spectral subplot. The \cii observations (in orange contours) clearly show the limited spectral resolution of FIFI-LS since each component is elongated along the velocity axis. Nevertheless, there are two regions where two components are clearly visible. The first one is between 0 and -16~arcsec south of the nucleus, a region which happens to be bright also in the near-UV. The second one is between 24 and 37~arcsec north of the nucleus in the region which has remnants of a past minor merger, as discussed in Section~\ref{sec:merging}.  The CO emission (green contours) shows at least two components in each section. The velocity dispersion in the nucleus (between 9 and -8 arcsec) is much higher than the spectral resolution. The weaker component, far from the major axis of the bar, has also a lower velocity dispersion than the component on the bar. Finally, we notice that there is a component of the UV emission north and south of the nucleus which is displaced with respect to the position of the bar. In the northern part the spiral arms are separated into two branches, as visible in the near-IR twin peaks. The brightest UV emission comes from the branch to the west of the bar, a location which precedes the bar in the sense of the galaxy rotation. The major axis \cii and CO emissions peak at the same location that also coincides with the peak of the UV emission. In the southern part of the bar, the situation is symmetric. A ridge of UV emission is visible east of the bar, again preceding the bar in the sense of galaxy rotation, but in this case there is no secondary branch in the infrared. The peak of the \cii emission is shifted towards east with respect to the CO emission below -8~arseconds south of the nucleus. There is some H$\alpha$ emission associated with the same region. The fact that \ion{H}{2} regions (outlined by the UV ridges) are displaced with respect to the position of the molecular gas has been also observed in other galaxy bars~\citep{Kartik2002}.
\begin{figure*}[!hbt]
\begin{center}
\includegraphics[width=0.7\textwidth]{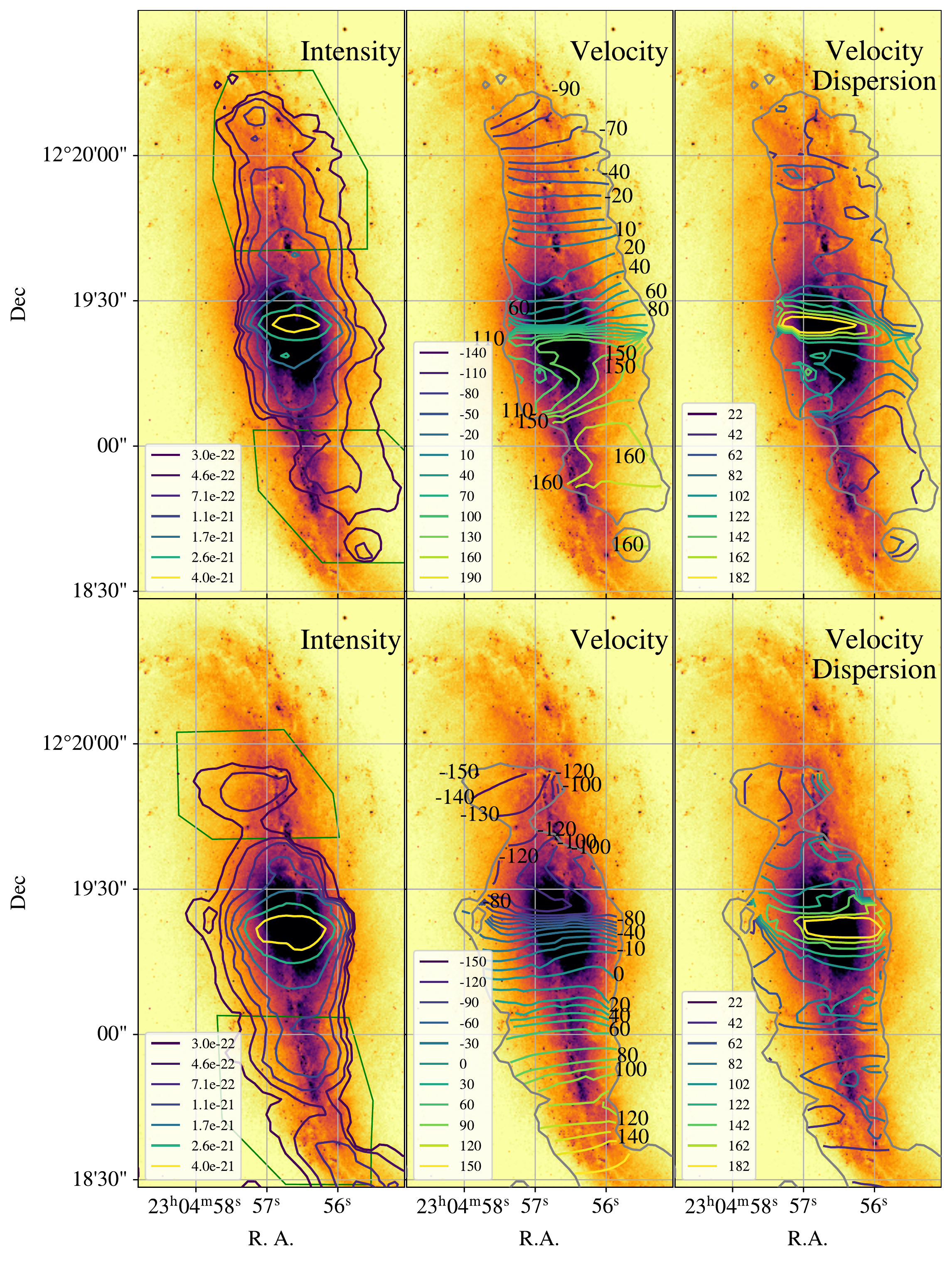}
\end{center}
\caption{The CO spectral cube separated into two kinematic components. This figure shows the logarithmic contours of the integrated emission, velocity, and velocity dispersion, for the  two components over the {\it HST} image of the galaxy. The top and bottom panels show the high and low velocity components, respectively. Levels are in W/m$^2$/pixel for intensity, and in km/s for velocity and velocity dispersion. The green polygons on the intensity plots are the regions we considered when estimating the amount of molecular gas along and outside the bar.
}
\label{fig:fields}
\end{figure*}

For each declination we identified the main components of the gas emission and fitted them with 2D~Gaussians. Fig.~\ref{fig:velra} shows the position and velocity of the peak of each component. The dispersion in velocity and R.A. is shown with horizontal bars. In the same figure (left panels) we report the location of the peak of the far-IR emission (at 70~$\mu$m) that accurately traces the location of the dust lane of the bar with a thick light blue line. Thin purple lines identify the locations where the UV emission peaks. The bulk of the molecular gas traced by the \cii and CO emission is found along the locations traced by the far-IR emission. The atomic gas traced by H$\alpha$ shows some emission in the outer regions. In particular, it traces the ridge of UV emission in the middle of the bar better than CO and \cii emissions. 

In the velocity plots (right panels), we traced a light-blue line with a slope of 5.6 km~s$^{-1}$~arcsec$^{-1}$ which marks the rotational velocity of the gas along the bar dust lanes. To identify the spatial structures in the velocity plot, components more than 4~arcsec from the major axis of the bar are plotted with a darker color. The CO velocity plot shows that the component along the bar has the typical profile of a galaxy bar that is similar to a rigid rotator. In the nuclear region (between $-15$ and $+15$~arcsec), the fast rotating nuclear disk \citep{1999ApJ...511..709L} distorts the velocity profile. Finally, the two fainter components in darker green appear to slow down until they reach the bar speed as they go out of the nucleus. The velocity plot of H$\alpha$ contains more components located outside of the bar. Nevertheless, all these components fall on the same pattern traced by the CO components. In particular, we notice that the UV ridges have some H$\alpha$ emission but their velocities show that these regions are linked to the bar. Finally, the \cii velocity plot shows two regions with two peaks: the southern part of the nucleus and the merging region. The less disturbed parts (ends of the bar) are again aligned along the velocity of the bar.

\subsection{An Interpretation of the Molecular Gas Flow}

The structure of the CO emission can be visualized in three~dimensions to better show the distribution of the gas along the bar. In Fig.~\ref{fig:tomography} we show the iso-contours of the CO emission for several sections in declination along the bar. Practically at each declination it is possible to distinguish at least two components distinct in velocity and with slightly different spatial locations. The strongest component is aligned with the bar, as shown in Fig.~\ref{fig:velra}. The weaker component detaches itself from the nucleus to join again the bar at its two ends. Because of the substantial difference in velocity of these two components with respect to the spectral resolution of the ALMA observations, it is possible to fit two lines for each spatial pixel in the CO spectral cube. In this way, we are able to separate the emission from the two velocity components.  Fig.~\ref{fig:fields} shows the integrated emission, velocity, and velocity dispersion of the two CO components. It is evident that the higher velocity component, in the top panels, includes most of the gas funnelled by the bar toward the galactic nucleus. The iso-velocity lines north of the nucleus are perpendicular to the dust lane, there is a steep gradient across the nucleus, and then they split in two diverging directions. Finally, south of the nucleus, the emission is mainly outside of the bar in a position trailing the rotation of the galaxy. A symmetrical situation is visible in the lower velocity component. This time, most of the emission is associated with the southern part of the bar, while north of the bar there is an accumulation of gas that seems to be limited by the horizontal dust lane in the bar. Again, the pattern of lines in the velocity fields is the same as in the northern component. We notice that the region where gas trails the bar is more extended in the low velocity component and also that there are several dust lanes associated with it, although not as sharp as the horizontal dust lane in the northern part of the bar.

It is natural to think that most of the gas in the stronger two velocity components on the opposite sides of the nucleus is simply gas flowing along the leading dust lanes of the bar towards the nucleus. This part exhibits the usual behavior of rigid rotation found in most barred galaxies, as the inflow velocity component is presumably much smaller than the component reflecting the tumbling of the bar. The weaker gas component, offset in velocity and trailing the bar, indicates that some gas may be falling into the bar, thus having a velocity component corresponding to the rotation curve of the galaxy in addition to the velocity component due to the participation in the tumbling of the bar. The origin of this gas is not clear. Since this structure has not been seen in any previous CO observations of galaxy bars, a possibility exists that such gas is linked to the minor merger, and corresponds at least partly to the ``anomalous dust lanes'' that intersect the bar almost at right angles. If the merging companion galaxy's orbit is almost perpendicular to the bar, some molecular gas from the disrupted galaxy could have escaped the bar, and rotated around the central part of the galaxy to be eventually recaptured by the bar in a later phase of the minor merger. In addition, some of this trailing gas forms naturally in a bar forming process (that may have been triggered by the minor merger), as seen in Figure 3 (left) of \citet{Laine1998b}, who simulated the bar pattern speed in NGC~7479 by matching the gas and dust morphology to the observations. We estimated the relative quantity of cold molecular gas trailing the bar by measuring the CO flux in the apertures drawn in Fig.~\ref{fig:fields}. The apertures are traced far enough from the nucleus to avoid contamination from the nuclear emission. The fluxes in the apertures on the bar are $(42 \pm 2)$~Jy~km/s north of the nucleus and $(31 \pm 3)$~Jy~km/s south of the nucleus. The fluxes in the regions beyond the nucleus are $(18.3\pm 0.3)$~Jy~km/s in the north and $(10\pm 1)$~Jy~km/s in the south. So, considering the total amount of gas along the bar, the ratio between the gas trailing the bar and the gas flowing towards the nucleus is around 40\%. Therefore, there is a non-negligible amount of gas that is not directly flowing along the bar. In particular, the cloud north of the nucleus located close to the horizontal dust lane has a flux of $(8.6 \pm 0.3)$~Jy~km/s. By using the relation between CO luminosity and molecular mass from \citet[][equation 3]{Bolatto2013} and a luminosity distance of 34.2~Mpc (assuming cz$=2381$~km/s), the mass of the cloud is approximately $10^8\ M_{\odot}$, bigger than a giant molecular cloud. However this is clearly not a stable cloud but it is probably formed by gas channeled to the bar through the horizontal dust lane. 

Refined simulations of the minor merging in NGC~7479, including gas and extending the work of \citet{Laine1999}, are needed to shed light on the mechanism responsible for the complex kinematics of the CO emission in this galaxy, but they are beyond the scope of the present paper.

\section{Summary and Conclusions} \label{sec:summary}

We presented the analysis of new SOFIA and ALMA observations of the [CII] and CO emission from the bar of the spiral galaxy NGC~7479. These observations have been compared to a wealth of archival photometric and spectroscopic observations, including unpublished Chandra observations. The main conclusions of this work can be summarized as follows.
\begin{itemize}
    \item We confirm the nuclear origin of the S-like structure found by \citet{Laine1998} by showing that the X-ray emission follows the same pattern as the 20~cm radio continuum. The X-ray observations confirm that the galaxy harbors a Compton-thick active nucleus. The spectrum extracted in the nuclear region has a high hardness ratio, and a broad Fe~K-alpha line at 6.4~keV is detected. The X-ray flux in the southern hot-spot exceeds the possible emission from pure star forming regions.
    \item Most of the \cii emission corresponds to CO emission in the bar showing that the majority of the \cii emission is due to cooling of gas excited through photoelectric heating by emission of young stars in PDRs. There are a few exceptions. At the ends of the X-ray/radio jet-like structure, the \cii emission is higher than what expected from the CO emission. However, infrared diagnostics show that the \cii emission in the northern end (region E) is mainly compatible with star formation. On the contrary, the southern end (region I) has an excess of \cii emission unrelated to star formation. We attribute this excess to the cooling of molecular gas shocked by the jet. Another location along the bar (region C) and one location external to the bar (region K) have very low CO/\cii ratios. These could be locations of CO-dark molecular clouds. Region K appears to have a low metallicity, and is the best candidate for CO-dark molecular gas.
    \item The high spectral resolution and sensitivity of the CO observations allowed us to separate the CO emission into two distinct kinematic components. Each velocity component consists of strong emission along the bar on one side with respect to the nucleus and of weak emission on the other side that trails the bar in the sense of the rotation of the galaxy. The gas trailing the bar is approximately 40\% of the gas along the bar, excluding the emission around the nucleus. In particular, a large cloud of molecular gas (mass of approximately $10^8$~M$_{\odot}$) is found on the location of a thick dust lane crossing the bar north of the nucleus, a feature probably related to a past minor merger. The origin of the gas trailing the bar is not clear. It could be related to the proposed minor merger in NGC~7479 where the companion has been mostly disrupted. 
\end{itemize}
A higher spectral resolution for the \cii observations would allow the separation of different kinematic components along the bar, thus enabling a better comparison with the ALMA CO data. Future observations of NGC~7479 with the GREAT spectrograph on SOFIA that has a spectral resolution similar to the existent ALMA observations are planned.

\acknowledgments
We thank Lauranne Lanz and Isaac Shlosman for illuminating discussions and suggestions. This research is based on data and software from the following projects. The NASA/DLR Stratospheric Observatory for Infrared Astronomy (SOFIA) jointly operated by USRA, under NASA contract NNA17BF53C, and DSI under DLR contract 50 OK 0901 to the University of Stuttgart. The ALMA observatory, operated by ESO, AUI/NRAO, and NAOJ, is a partnership of ESO, NSF (USA), and NINS (Japan), together with NRC (Canada), MOST and ASIAA (Taiwan), and KASI (Rep. of Korea), in cooperation with the Rep. of Chile. Herschel, an ESA space observatory science with instruments provided by European-led P.I. consortia and important NASA participation. The Spitzer Space Telescope, operated by JPL, Caltech under a contract with NASA. The SDSS survey, funded by the A. P. Sloan Foundation, the Participating Institutions, NSF, the U.S. Dep. of Energy, NASA, the Japanese Monbukagakusho, the Max Planck Society, and the Higher Education Funding Council for England. The Two Micron All Sky Survey (2MASS), a joint project of the University of Massachusetts and IPAC/Caltech, funded by NASA and NSF. The GALEX archive hosted by the High Energy Astrophysics Science Archive Research Center (HEASARC), which is a service of the Astrophysics Science Division at NASA/GSFC. The Chandra Data Archive and the CIAO software provided by the Chandra X-ray Center. Financial support for the project was provided by NASA through award \# SOF-06-0124 issued by USRA.


%\vspace{5mm}
\facilities{SOFIA (FIFI-LS), Spitzer (IRAC, MIPS), Herschel (PACS, SPIRE), ALMA, GALEX, Chandra, SDSS, 2MASS}

\software{
astropy \citep{2013A&A...558A..33A}, 
sospex (\citet{2018AAS...23115011F}, \url{http://www.github.com/darioflute/sospex}), mopex (\citet{2005PASP..117.1113M}, \url{https://irsa.ipac.caltech.edu/data/SPITZER/docs/dataanalysistools/tools/mopex/}),
stinytim (John Krist,\url{https://irsa.ipac.caltech.edu/data/SPITZER/docs/dataanalysistools/tools/contributed/general/stinytim/}),
CIAO (\citet{Fruscione2006}, \url{https://cxc.harvard.edu/ciao4.12/}),
magphys (\citet{daCunha2008},  \url{http://www.iap.fr/magphys/})
}

\newpage
\bibliography{biblio}{}

\begin{thebibliography}{}
\expandafter\ifx\csname natexlab\endcsname\relax\def\natexlab#1{#1}\fi
\providecommand{\url}[1]{\href{#1}{#1}}
\providecommand{\dodoi}[1]{doi:~\href{http://doi.org/#1}{\nolinkurl{#1}}}
\providecommand{\doeprint}[1]{\href{http://ascl.net/#1}{\nolinkurl{http://ascl.net/#1}}}
\providecommand{\doarXiv}[1]{\href{https://arxiv.org/abs/#1}{\nolinkurl{https://arxiv.org/abs/#1}}}

\bibitem[{{Appleton} {et~al.}(2013){Appleton}, {Guillard}, {Boulanger},
  {Cluver}, {Ogle}, {Falgarone}, {Pineau des For{\^e}ts}, {O'Sullivan}, {Duc},
  {Gallagher}, {Gao}, {Jarrett}, {Konstantopoulos}, {Lisenfeld}, {Lord}, {Lu},
  {Peterson}, {Struck}, {Sturm}, {Tuffs}, {Valchanov}, {van der Werf}, \&
  {Xu}}]{Appleton2013}
{Appleton}, P.~N., {Guillard}, P., {Boulanger}, F., {et~al.} 2013, \apj, 777,
  66, \dodoi{10.1088/0004-637X/777/1/66}

\bibitem[{{Appleton} {et~al.}(2017){Appleton}, {Guillard}, {Togi}, {Alatalo},
  {Boulanger}, {Cluver}, {Pineau des For{\^e}ts}, {Lisenfeld}, {Ogle}, \&
  {Xu}}]{Appleton2017}
{Appleton}, P.~N., {Guillard}, P., {Togi}, A., {et~al.} 2017, \apj, 836, 76,
  \dodoi{10.3847/1538-4357/836/1/76}

\bibitem[{{Appleton} {et~al.}(2018){Appleton}, {Diaz-Santos}, {Fadda}, {Ogle},
  {Togi}, {Lanz}, {Alatalo}, {Fischer}, {Rich}, \& {Guillard}}]{Appleton2018}
{Appleton}, P.~N., {Diaz-Santos}, T., {Fadda}, D., {et~al.} 2018, \apj, 869,
  61, \dodoi{10.3847/1538-4357/aaed2a}

\bibitem[{{Astropy Collaboration} {et~al.}(2013){Astropy Collaboration},
  {Robitaille}, {Tollerud}, {Greenfield}, {Droettboom}, {Bray}, {Aldcroft},
  {Davis}, {Ginsburg}, {Price-Whelan}, {Kerzendorf}, {Conley}, {Crighton},
  {Barbary}, {Muna}, {Ferguson}, {Grollier}, {Parikh}, {Nair}, {Unther},
  {Deil}, {Woillez}, {Conseil}, {Kramer}, {Turner}, {Singer}, {Fox}, {Weaver},
  {Zabalza}, {Edwards}, {Azalee Bostroem}, {Burke}, {Casey}, {Crawford},
  {Dencheva}, {Ely}, {Jenness}, {Labrie}, {Lim}, {Pierfederici}, {Pontzen},
  {Ptak}, {Refsdal}, {Servillat}, \& {Streicher}}]{2013A&A...558A..33A}
{Astropy Collaboration}, {Robitaille}, T.~P., {Tollerud}, E.~J., {et~al.} 2013,
  \aap, 558, A33, \dodoi{10.1051/0004-6361/201322068}

\bibitem[{{Bakes} \& {Tielens}(1998)}]{Bakes1998}
{Bakes}, E.~L.~O., \& {Tielens}, A.~G.~G.~M. 1998, \apj, 499, 258,
  \dodoi{10.1086/305625}

\bibitem[{{Barnes} \& {Hernquist}(1991)}]{Barnes1991}
{Barnes}, J.~E., \& {Hernquist}, L.~E. 1991, \apjl, 370, L65,
  \dodoi{10.1086/185978}

\bibitem[{{Beir{\~a}o} {et~al.}(2012){Beir{\~a}o}, {Armus}, {Helou},
  {Appleton}, {Smith}, {Croxall}, {Murphy}, {Dale}, {Draine}, {Wolfire}, {Sand
  strom}, {Aniano}, {Bolatto}, {Groves}, {Brandl}, {Schinnerer}, {Crocker},
  {Hinz}, {Rix}, {Kennicutt}, {Calzetti}, {Gil de Paz}, {Dumas}, {Galametz},
  {Gordon}, {Hao}, {Johnson}, {Koda}, {Krause}, {van der Laan}, {Leroy}, {Li},
  {Meidt}, {Meyer}, {Rahman}, {Roussel}, {Sauvage}, {Srinivasan}, {Vigroux},
  {Walter}, \& {Warren}}]{Beirao2012}
{Beir{\~a}o}, P., {Armus}, L., {Helou}, G., {et~al.} 2012, \apj, 751, 144,
  \dodoi{10.1088/0004-637X/751/2/144}

\bibitem[{{Bigiel} {et~al.}(2020){Bigiel}, {de Looze}, {Krabbe}, {Cormier},
  {Barnes}, {Fischer}, {Bolatto}, {Bryant}, {Colditz}, {Geis}, {Herrera-Camus},
  {Iserlohe}, {Klein}, {Leroy}, {Linz}, {Looney}, {Madden}, {Poglitsch},
  {Stutzki}, \& {Vacca}}]{Bigiel2020}
{Bigiel}, F., {de Looze}, I., {Krabbe}, A., {et~al.} 2020, \apj, 903, 30,
  \dodoi{10.3847/1538-4357/abb677}

\bibitem[{{Blumenthal} \& {Barnes}(2018)}]{Blumenthal2018}
{Blumenthal}, K.~A., \& {Barnes}, J.~E. 2018, \mnras, 479, 3952,
  \dodoi{10.1093/mnras/sty1605}

\bibitem[{{Bolatto} {et~al.}(2013){Bolatto}, {Wolfire}, \&
  {Leroy}}]{Bolatto2013}
{Bolatto}, A.~D., {Wolfire}, M., \& {Leroy}, A.~K. 2013, \araa, 51, 207,
  \dodoi{10.1146/annurev-astro-082812-140944}

\bibitem[{{Burtscher} {et~al.}(2016){Burtscher}, {Davies}, {Graci{\'a}-Carpio},
  {Koss}, {Lin}, {Lutz}, {Nandra}, {Netzer}, {Orban de Xivry}, {Ricci},
  {Rosario}, {Veilleux}, {Contursi}, {Genzel}, {Schnorr-M{\"u}ller},
  {Sternberg}, {Sturm}, \& {Tacconi}}]{Burtscher2016}
{Burtscher}, L., {Davies}, R.~I., {Graci{\'a}-Carpio}, J., {et~al.} 2016, \aap,
  586, A28, \dodoi{10.1051/0004-6361/201527575}

\bibitem[{{Chevance} {et~al.}(2020){Chevance}, {Madden}, {Fischer}, {Vacca},
  {Lebouteiller}, {Fadda}, {Galliano}, {Indebetouw}, {Kruijssen}, {Lee},
  {Poglitsch}, {Polles}, {Cormier}, {Hony}, {Iserlohe}, {Krabbe}, {Meixner},
  {Sabbi}, \& {Zinnecker}}]{Chevance2020}
{Chevance}, M., {Madden}, S.~C., {Fischer}, C., {et~al.} 2020, \mnras, 494,
  5279, \dodoi{10.1093/mnras/staa1106}

\bibitem[{{Colditz} {et~al.}(2018){Colditz}, {Beckmann}, {Bryant}, {Fischer},
  {Fumi}, {Geis}, {Hamidouche}, {Henning}, {H{\"o}nle}, {Iserlohe}, {Klein},
  {Krabbe}, {Looney}, {Poglitsch}, {Raab}, {Rebell}, {Rosenthal}, {Savage},
  {Schweitzer}, \& {Vacca}}]{2018JAI.....740004C}
{Colditz}, S., {Beckmann}, S., {Bryant}, A., {et~al.} 2018, Journal of
  Astronomical Instrumentation, 7, 1840004, \dodoi{10.1142/S2251171718400044}

\bibitem[{{Croxall} {et~al.}(2012){Croxall}, {Smith}, {Wolfire}, {Roussel},
  {Sandstrom}, {Draine}, {Aniano}, {Dale}, {Armus}, {Beir{\~a}o}, {Helou},
  {Bolatto}, {Appleton}, {Brand l}, {Calzetti}, {Crocker}, {Galametz},
  {Groves}, {Hao}, {Hunt}, {Johnson}, {Kennicutt}, {Koda}, {Krause}, {Li},
  {Meidt}, {Murphy}, {Rahman}, {Rix}, {Sauvage}, {Schinnerer}, {Walter}, \&
  {Wilson}}]{Croxall2012}
{Croxall}, K.~V., {Smith}, J.~D., {Wolfire}, M.~G., {et~al.} 2012, \apj, 747,
  81, \dodoi{10.1088/0004-637X/747/1/81}

\bibitem[{{Croxall} {et~al.}(2017){Croxall}, {Smith}, {Pellegrini}, {Groves},
  {Bolatto}, {Herrera-Camus}, {Sand strom}, {Draine}, {Wolfire}, {Armus},
  {Boquien}, {Brandl}, {Dale}, {Galametz}, {Hunt}, {Kennicutt}, {Kreckel},
  {Rigopoulou}, {van der Werf}, \& {Wilson}}]{Croxall2017}
{Croxall}, K.~V., {Smith}, J.~D., {Pellegrini}, E., {et~al.} 2017, \apj, 845,
  96, \dodoi{10.3847/1538-4357/aa8035}

\bibitem[{{da Cunha} {et~al.}(2008){da Cunha}, {Charlot}, \&
  {Elbaz}}]{daCunha2008}
{da Cunha}, E., {Charlot}, S., \& {Elbaz}, D. 2008, \mnras, 388, 1595,
  \dodoi{10.1111/j.1365-2966.2008.13535.x}

\bibitem[{{Daigle} \& {Roy}(2001)}]{Daigle2001}
{Daigle}, A., \& {Roy}, J.-R. 2001, \apj, 552, 144, \dodoi{10.1086/320437}

\bibitem[{{De Looze} {et~al.}(2011){De Looze}, {Baes}, {Bendo}, {Cortese}, \&
  {Fritz}}]{DeLooze2011}
{De Looze}, I., {Baes}, M., {Bendo}, G.~J., {Cortese}, L., \& {Fritz}, J. 2011,
  \mnras, 416, 2712, \dodoi{10.1111/j.1365-2966.2011.19223.x}

\bibitem[{{D{\'\i}az--Santos} {et~al.}(2014){D{\'\i}az--Santos}, {Armus},
  {Charmandaris}, {Stacey}, {Murphy}, {Haan}, {Stierwalt}, {Malhotra},
  {Appleton}, {Inami}, {Magdis}, {Elbaz}, {Evans}, {Mazzarella}, {Surace}, {van
  der Werf}, {Xu}, {Lu}, {Meijerink}, {Howell}, {Petric}, {Veilleux}, \&
  {Sanders}}]{Diaz-Santos2014}
{D{\'\i}az--Santos}, T., {Armus}, L., {Charmandaris}, V., {et~al.} 2014, \apjl,
  788, L17, \dodoi{10.1088/2041-8205/788/1/L17}

\bibitem[{{D{\'\i}az--Santos} {et~al.}(2017){D{\'\i}az--Santos}, {Armus},
  {Charmandaris}, {Lu}, {Stierwalt}, {Stacey}, {Malhotra}, {van der Werf},
  {Howell}, {Privon}, {Mazzarella}, {Goldsmith}, {Murphy}, {Barcos-Mu{\~n}oz},
  {Linden}, {Inami}, {Larson}, {Evans}, {Appleton}, {Iwasawa}, {Lord},
  {Sanders}, \& {Surace}}]{DiazSantos2017}
---. 2017, \apj, 846, 32, \dodoi{10.3847/1538-4357/aa81d7}

\bibitem[{{Draine}(1978)}]{Draine1978}
{Draine}, B.~T. 1978, \apjs, 36, 595, \dodoi{10.1086/190513}

\bibitem[{{Eskridge} {et~al.}(2000){Eskridge}, {Frogel}, {Pogge}, {Quillen},
  {Davies}, {DePoy}, {Houdashelt}, {Kuchinski}, {Ram{\'\i}rez}, {Sellgren},
  {Terndrup}, \& {Tiede}}]{Eskridge2000}
{Eskridge}, P.~B., {Frogel}, J.~A., {Pogge}, R.~W., {et~al.} 2000, \aj, 119,
  536, \dodoi{10.1086/301203}

\bibitem[{{Fadda} \& {Chambers}(2018)}]{2018AAS...23115011F}
{Fadda}, D., \& {Chambers}, E.~T. 2018, in American Astronomical Society
  Meeting Abstracts, Vol. 231, American Astronomical Society Meeting Abstracts
  \#231, 150.11

\bibitem[{{Fadda} {et~al.}(2006){Fadda}, {Marleau}, {Storrie-Lombardi},
  {Makovoz}, {Frayer}, {Appleton}, {Armus}, {Chapman}, {Choi}, {Fang},
  {Heinrichsen}, {Helou}, {Im}, {Lacy}, {Shupe}, {Soifer}, {Squires}, {Surace},
  {Teplitz}, {Wilson}, \& {Yan}}]{Fadda2006}
{Fadda}, D., {Marleau}, F.~R., {Storrie-Lombardi}, L.~J., {et~al.} 2006, \aj,
  131, 2859, \dodoi{10.1086/504034}

\bibitem[{{Fazio} {et~al.}(2004){Fazio}, {Hora}, {Allen}, {Ashby}, {Barmby},
  {Deutsch}, {Huang}, {Kleiner}, {Marengo}, {Megeath}, {Melnick}, {Pahre},
  {Patten}, {Polizotti}, {Smith}, {Taylor}, {Wang}, {Willner}, {Hoffmann},
  {Pipher}, {Forrest}, {McMurty}, {McCreight}, {McKelvey}, {McMurray}, {Koch},
  {Moseley}, {Arendt}, {Mentzell}, {Marx}, {Losch}, {Mayman}, {Eichhorn},
  {Krebs}, {Jhabvala}, {Gezari}, {Fixsen}, {Flores}, {Shakoorzadeh}, {Jungo},
  {Hakun}, {Workman}, {Karpati}, {Kichak}, {Whitley}, {Mann}, {Tollestrup},
  {Eisenhardt}, {Stern}, {Gorjian}, {Bhattacharya}, {Carey}, {Nelson},
  {Glaccum}, {Lacy}, {Lowrance}, {Laine}, {Reach}, {Stauffer}, {Surace},
  {Wilson}, {Wright}, {Hoffman}, {Domingo}, \& {Cohen}}]{2004ApJS..154...10F}
{Fazio}, G.~G., {Hora}, J.~L., {Allen}, L.~E., {et~al.} 2004, \apjs, 154, 10,
  \dodoi{10.1086/422843}

\bibitem[{{Fischer} {et~al.}(2018){Fischer}, {Beckmann}, {Bryant}, {Colditz},
  {Fumi}, {Geis}, {Hamidouche}, {Henning}, {H{\"o}nle}, {Iserlohe}, {Klein},
  {Krabbe}, {Looney}, {Poglitsch}, {Raab}, {Rebell}, {Rosenthal}, {Savage},
  {Schweitzer}, {Trinh}, \& {Vacca}}]{2018JAI.....740003F}
{Fischer}, C., {Beckmann}, S., {Bryant}, A., {et~al.} 2018, Journal of
  Astronomical Instrumentation, 7, 1840003, \dodoi{10.1142/S2251171718400032}

\bibitem[{{Fruscione} {et~al.}(2006){Fruscione}, {McDowell}, {Allen},
  {Brickhouse}, {Burke}, {Davis}, {Durham}, {Elvis}, {Galle}, {Harris},
  {Huenemoerder}, {Houck}, {Ishibashi}, {Karovska}, {Nicastro}, {Noble},
  {Nowak}, {Primini}, {Siemiginowska}, {Smith}, \& {Wise}}]{Fruscione2006}
{Fruscione}, A., {McDowell}, J.~C., {Allen}, G.~E., {et~al.} 2006, in Society
  of Photo-Optical Instrumentation Engineers (SPIE) Conference Series, Vol.
  6270, Society of Photo-Optical Instrumentation Engineers (SPIE) Conference
  Series, ed. D.~R. {Silva} \& R.~E. {Doxsey}, 62701V,
  \dodoi{10.1117/12.671760}

\bibitem[{{Guillard} {et~al.}(2009){Guillard}, {Boulanger}, {Pineau Des
  For{\^e}ts}, \& {Appleton}}]{Guillard2009}
{Guillard}, P., {Boulanger}, F., {Pineau Des For{\^e}ts}, G., \& {Appleton},
  P.~N. 2009, \aap, 502, 515, \dodoi{10.1051/0004-6361/200811263}

\bibitem[{{Herrera--Camus} {et~al.}(2015){Herrera--Camus}, {Bolatto},
  {Wolfire}, {Smith}, {Croxall}, {Kennicutt}, {Calzetti}, {Helou}, {Walter},
  {Leroy}, {Draine}, {Brandl}, {Armus}, {Sand strom}, {Dale}, {Aniano},
  {Meidt}, {Boquien}, {Hunt}, {Galametz}, {Tabatabaei}, {Murphy}, {Appleton},
  {Roussel}, {Engelbracht}, \& {Beirao}}]{Herrera2015}
{Herrera--Camus}, R., {Bolatto}, A.~D., {Wolfire}, M.~G., {et~al.} 2015, \apj,
  800, 1, \dodoi{10.1088/0004-637X/800/1/1}

\bibitem[{{Ho} {et~al.}(1997){Ho}, {Filippenko}, \& {Sargent}}]{Ho1997}
{Ho}, L.~C., {Filippenko}, A.~V., \& {Sargent}, W. L.~W. 1997, \apj, 487, 591,
  \dodoi{10.1086/304643}

\bibitem[{{Hollenbach} \& {McKee}(1989)}]{Hollenbach1989}
{Hollenbach}, D., \& {McKee}, C.~F. 1989, \apj, 342, 306,
  \dodoi{10.1086/167595}

\bibitem[{{Ishibashi} \& {Fabian}(2012)}]{Ishibashi2012}
{Ishibashi}, W., \& {Fabian}, A.~C. 2012, \mnras, 427, 2998,
  \dodoi{10.1111/j.1365-2966.2012.22074.x}

\bibitem[{{Jameson} {et~al.}(2018){Jameson}, {Bolatto}, {Wolfire}, {Warren},
  {Herrera-Camus}, {Croxall}, {Pellegrini}, {Smith}, {Rubio}, {Indebetouw},
  {Israel}, {Meixner}, {Roman-Duval}, {van Loon}, {Muller}, {Verdugo},
  {Zinnecker}, \& {Okada}}]{Jameson2018}
{Jameson}, K.~E., {Bolatto}, A.~D., {Wolfire}, M., {et~al.} 2018, \apj, 853,
  111, \dodoi{10.3847/1538-4357/aaa4bb}

\bibitem[{{Jogee} {et~al.}(2009){Jogee}, {Miller}, {Penner}, {Skelton},
  {Conselice}, {Somerville}, {Bell}, {Zheng}, {Rix}, {Robaina}, {Barazza},
  {Barden}, {Borch}, {Beckwith}, {Caldwell}, {Peng}, {Heymans}, {McIntosh},
  {H{\"a}u{\ss}ler}, {Jahnke}, {Meisenheimer}, {Sanchez}, {Wisotzki}, {Wolf},
  \& {Papovich}}]{Jogee2009}
{Jogee}, S., {Miller}, S.~H., {Penner}, K., {et~al.} 2009, \apj, 697, 1971,
  \dodoi{10.1088/0004-637X/697/2/1971}

\bibitem[{{Kaufman} {et~al.}(1999){Kaufman}, {Wolfire}, {Hollenbach}, \&
  {Luhman}}]{Kaufman1999}
{Kaufman}, M.~J., {Wolfire}, M.~G., {Hollenbach}, D.~J., \& {Luhman}, M.~L.
  1999, \apj, 527, 795, \dodoi{10.1086/308102}

\bibitem[{{Kaviraj}(2014)}]{Kaviraj2014}
{Kaviraj}, S. 2014, \mnras, 440, 2944, \dodoi{10.1093/mnras/stu338}

\bibitem[{{Keel}(1983)}]{Keel1983}
{Keel}, W.~C. 1983, \apjs, 52, 229, \dodoi{10.1086/190866}

\bibitem[{{Kendall} {et~al.}(2003){Kendall}, {Magorrian}, \&
  {Pringle}}]{Kendall2003}
{Kendall}, P., {Magorrian}, J., \& {Pringle}, J.~E. 2003, \mnras, 346, 1078,
  \dodoi{10.1111/j.1365-2966.2003.06776.x}

\bibitem[{{Kennicutt}(1998)}]{Kennicutt1998}
{Kennicutt}, Robert~C., J. 1998, \apj, 498, 541, \dodoi{10.1086/305588}

\bibitem[{{Kewley} {et~al.}(2006){Kewley}, {Groves}, {Kauffmann}, \&
  {Heckman}}]{Kewley2006}
{Kewley}, L.~J., {Groves}, B., {Kauffmann}, G., \& {Heckman}, T. 2006, \mnras,
  372, 961, \dodoi{10.1111/j.1365-2966.2006.10859.x}

\bibitem[{{Laine} \& {Beck}(2008)}]{Laine2008}
{Laine}, S., \& {Beck}, R. 2008, \apj, 673, 128, \dodoi{10.1086/523960}

\bibitem[{{Laine} \& {Gottesman}(1998)}]{Laine1998}
{Laine}, S., \& {Gottesman}, S.~T. 1998, \mnras, 297, 1041,
  \dodoi{10.1046/j.1365-8711.1998.01513.x}

\bibitem[{{Laine} \& {Heller}(1999)}]{Laine1999}
{Laine}, S., \& {Heller}, C.~H. 1999, \mnras, 308, 557,
  \dodoi{10.1046/j.1365-8711.1999.02712.x}

\bibitem[{{Laine} {et~al.}(1999){Laine}, {Kenney}, {Yun}, \&
  {Gottesman}}]{1999ApJ...511..709L}
{Laine}, S., {Kenney}, J.~D.~P., {Yun}, M.~S., \& {Gottesman}, S.~T. 1999,
  \apj, 511, 709, \dodoi{10.1086/306709}

\bibitem[{{Laine} {et~al.}(1998){Laine}, {Shlosman}, \& {Heller}}]{Laine1998b}
{Laine}, S., {Shlosman}, I., \& {Heller}, C.~H. 1998, \mnras, 297, 1052,
  \dodoi{10.1046/j.1365-8711.1998.01512.x}

\bibitem[{{Lesaffre} {et~al.}(2013){Lesaffre}, {Pineau des For{\^e}ts},
  {Godard}, {Guillard}, {Boulanger}, \& {Falgarone}}]{Lesaffre2013}
{Lesaffre}, P., {Pineau des For{\^e}ts}, G., {Godard}, B., {et~al.} 2013, \aap,
  550, A106, \dodoi{10.1051/0004-6361/201219928}

\bibitem[{{Lord}(1992)}]{1992nstc.rept.....L}
{Lord}, S.~D. 1992, {A new software tool for computing Earth's atmospheric
  transmission of near- and far-infrared radiation}, NASA Technical Memorandum
  103957

\bibitem[{{Madden} {et~al.}(2020){Madden}, {Cormier}, {Hony}, {Lebouteiller},
  {Abel}, {Galametz}, {De Looze}, {Chevance}, {Polles}, {Lee}, {Galliano},
  {Lambert-Huyghe}, {Hu}, \& {Ramambason}}]{Madden2020}
{Madden}, S.~C., {Cormier}, D., {Hony}, S., {et~al.} 2020, \aap, 643, A141,
  \dodoi{10.1051/0004-6361/202038860}

\bibitem[{{Maiolino} {et~al.}(1999){Maiolino}, {Risaliti}, \&
  {Salvati}}]{Maiolino1999}
{Maiolino}, R., {Risaliti}, G., \& {Salvati}, M. 1999, \aap, 341, L35.
\newblock \doarXiv{astro-ph/9811237}

\bibitem[{{Makovoz} \& {Marleau}(2005)}]{2005PASP..117.1113M}
{Makovoz}, D., \& {Marleau}, F.~R. 2005, \pasp, 117, 1113,
  \dodoi{10.1086/432977}

\bibitem[{{Malhotra} {et~al.}(2001){Malhotra}, {Kaufman}, {Hollenbach},
  {Helou}, {Rubin}, {Brauher}, {Dale}, {Lu}, {Lord}, {Stacey}, {Contursi},
  {Hunter}, \& {Dinerstein}}]{Malhotra2001}
{Malhotra}, S., {Kaufman}, M.~J., {Hollenbach}, D., {et~al.} 2001, \apj, 561,
  766, \dodoi{10.1086/323046}

\bibitem[{{Martin} {et~al.}(2000){Martin}, {Leli{\`e}vre}, \&
  {Roy}}]{Martin2000}
{Martin}, P., {Leli{\`e}vre}, M., \& {Roy}, J.-R. 2000, \apj, 538, 141,
  \dodoi{10.1086/309131}

\bibitem[{{Martin} \& {Roy}(1994)}]{Martin1994}
{Martin}, P., \& {Roy}, J.-R. 1994, \apj, 424, 599, \dodoi{10.1086/173917}

\bibitem[{{Masters} {et~al.}(2012){Masters}, {Nichol}, {Haynes}, {Keel},
  {Lintott}, {Simmons}, {Skibba}, {Bamford}, {Giovanelli}, \&
  {Schawinski}}]{Masters2012}
{Masters}, K.~L., {Nichol}, R.~C., {Haynes}, M.~P., {et~al.} 2012, \mnras, 424,
  2180, \dodoi{10.1111/j.1365-2966.2012.21377.x}

\bibitem[{{Mihos} \& {Hernquist}(1994)}]{Mihos1994}
{Mihos}, J.~C., \& {Hernquist}, L. 1994, \apjl, 425, L13,
  \dodoi{10.1086/187299}

\bibitem[{{Mineo} {et~al.}(2012){Mineo}, {Gilfanov}, \& {Sunyaev}}]{Mineo2012}
{Mineo}, S., {Gilfanov}, M., \& {Sunyaev}, R. 2012, \mnras, 419, 2095,
  \dodoi{10.1111/j.1365-2966.2011.19862.x}

\bibitem[{{Mukherjee} {et~al.}(2016){Mukherjee}, {Bicknell}, {Sutherland}, \&
  {Wagner}}]{Mukherjee2016}
{Mukherjee}, D., {Bicknell}, G.~V., {Sutherland}, R., \& {Wagner}, A. 2016,
  \mnras, 461, 967, \dodoi{10.1093/mnras/stw1368}

\bibitem[{{Peeters} {et~al.}(2004){Peeters}, {Spoon}, \&
  {Tielens}}]{Peters2004}
{Peeters}, E., {Spoon}, H.~W.~W., \& {Tielens}, A.~G.~G.~M. 2004, \apj, 613,
  986, \dodoi{10.1086/423237}

\bibitem[{{Peterson} {et~al.}(2018){Peterson}, {Appleton}, {Bitsakis},
  {Guillard}, {Alatalo}, {Boulanger}, {Cluver}, {Duc}, {Falgarone},
  {Gallagher}, {Gao}, {Helou}, {Jarrett}, {Joshi}, {Lisenfeld}, {Lu}, {Ogle},
  {Pineau des For{\^e}ts}, {van der Werf}, \& {Xu}}]{Peterson2018}
{Peterson}, B.~W., {Appleton}, P.~N., {Bitsakis}, T., {et~al.} 2018, \apj, 855,
  141, \dodoi{10.3847/1538-4357/aaac2c}

\bibitem[{{Pineda} {et~al.}(2013){Pineda}, {Langer}, {Velusamy}, \&
  {Goldsmith}}]{Pineda2013}
{Pineda}, J.~L., {Langer}, W.~D., {Velusamy}, T., \& {Goldsmith}, P.~F. 2013,
  \aap, 554, A103, \dodoi{10.1051/0004-6361/201321188}

\bibitem[{{Pineda} {et~al.}(2018){Pineda}, {Fischer}, {Kapala}, {Stutzki},
  {Buchbender}, {Goldsmith}, {Ziebart}, {Glover}, {Klessen}, {Koda}, {Kramer},
  {Mookerjea}, {Sandstrom}, {Scoville}, \& {Smith}}]{Pineda2018}
{Pineda}, J.~L., {Fischer}, C., {Kapala}, M., {et~al.} 2018, \apjl, 869, L30,
  \dodoi{10.3847/2041-8213/aaf1ad}

\bibitem[{{Plante} {et~al.}(1991){Plante}, {Lo}, {Roy}, {Martin}, \&
  {Noreau}}]{Plante1991}
{Plante}, R.~L., {Lo}, K.~Y., {Roy}, J.-R., {Martin}, P., \& {Noreau}, L. 1991,
  \apj, 381, 110, \dodoi{10.1086/170633}

\bibitem[{{Quillen} {et~al.}(1995){Quillen}, {Frogel}, {Kenney}, {Pogge}, \&
  {Depoy}}]{Quillen1995}
{Quillen}, A.~C., {Frogel}, J.~A., {Kenney}, J. D.~P., {Pogge}, R.~W., \&
  {Depoy}, D.~L. 1995, \apj, 441, 549, \dodoi{10.1086/175381}

\bibitem[{{Regan} {et~al.}(1999){Regan}, {Sheth}, \& {Vogel}}]{Regan1999}
{Regan}, M.~W., {Sheth}, K., \& {Vogel}, S.~N. 1999, \apj, 526, 97,
  \dodoi{10.1086/307960}

\bibitem[{{Rieke} {et~al.}(2004){Rieke}, {Young}, {Engelbracht}, {Kelly},
  {Low}, {Haller}, {Beeman}, {Gordon}, {Stansberry}, {Misselt}, {Cadien},
  {Morrison}, {Rivlis}, {Latter}, {Noriega-Crespo}, {Padgett}, {Stapelfeldt},
  {Hines}, {Egami}, {Muzerolle}, {Alonso-Herrero}, {Blaylock}, {Dole}, {Hinz},
  {Le Floc'h}, {Papovich}, {P{\'e}rez-Gonz{\'a}lez}, {Smith}, {Su}, {Bennett},
  {Frayer}, {Henderson}, {Lu}, {Masci}, {Pesenson}, {Rebull}, {Rho}, {Keene},
  {Stolovy}, {Wachter}, {Wheaton}, {Werner}, \&
  {Richards}}]{2004ApJS..154...25R}
{Rieke}, G.~H., {Young}, E.~T., {Engelbracht}, C.~W., {et~al.} 2004, \apjs,
  154, 25, \dodoi{10.1086/422717}

\bibitem[{{Sakamoto} {et~al.}(1999){Sakamoto}, {Okumura}, {Ishizuki}, \&
  {Scoville}}]{Sakamoto1999}
{Sakamoto}, K., {Okumura}, S.~K., {Ishizuki}, S., \& {Scoville}, N.~Z. 1999,
  \apj, 525, 691, \dodoi{10.1086/307910}

\bibitem[{{Schmidt} {et~al.}(2018){Schmidt}, {Oio}, {Ferreiro}, {Vega}, \&
  {Weidmann}}]{Schmidt2018}
{Schmidt}, E.~O., {Oio}, G.~A., {Ferreiro}, D., {Vega}, L., \& {Weidmann}, W.
  2018, \aap, 615, A13, \dodoi{10.1051/0004-6361/201731557}

\bibitem[{{Sheth} {et~al.}(2002){Sheth}, {Vogel}, {Regan}, {Teuben}, {Harris},
  \& {Thornley}}]{Kartik2002}
{Sheth}, K., {Vogel}, S.~N., {Regan}, M.~W., {et~al.} 2002, \aj, 124, 2581,
  \dodoi{10.1086/343835}

\bibitem[{{Sheth} {et~al.}(2005){Sheth}, {Vogel}, {Regan}, {Thornley}, \&
  {Teuben}}]{Kartik2005}
{Sheth}, K., {Vogel}, S.~N., {Regan}, M.~W., {Thornley}, M.~D., \& {Teuben},
  P.~J. 2005, \apj, 632, 217, \dodoi{10.1086/432409}

\bibitem[{{Sheth} {et~al.}(2008){Sheth}, {Elmegreen}, {Elmegreen}, {Capak},
  {Abraham}, {Athanassoula}, {Ellis}, {Mobasher}, {Salvato}, {Schinnerer},
  {Scoville}, {Spalsbury}, {Strubbe}, {Carollo}, {Rich}, \&
  {West}}]{Kartik2008}
{Sheth}, K., {Elmegreen}, D.~M., {Elmegreen}, B.~G., {et~al.} 2008, \apj, 675,
  1141, \dodoi{10.1086/524980}

\bibitem[{{Silk}(2013)}]{silk2013}
{Silk}, J. 2013, \apj, 772, 112, \dodoi{10.1088/0004-637X/772/2/112}

\bibitem[{{Stacey} {et~al.}(1991){Stacey}, {Geis}, {Genzel}, {Lugten},
  {Poglitsch}, {Sternberg}, \& {Townes}}]{Stacey1991}
{Stacey}, G.~J., {Geis}, N., {Genzel}, R., {et~al.} 1991, \apj, 373, 423,
  \dodoi{10.1086/170062}

\bibitem[{{Taniguchi}(1999)}]{taniguchi1999}
{Taniguchi}, Y. 1999, \apj, 524, 65, \dodoi{10.1086/307814}

\bibitem[{{Tielens} \& {Hollenbach}(1985)}]{Tielens1985}
{Tielens}, A.~G.~G.~M., \& {Hollenbach}, D. 1985, \apj, 291, 722,
  \dodoi{10.1086/163111}

\bibitem[{{Vacca}(2020)}]{2020ASPC..Vacca}
{Vacca}, W.~D. 2020, {The Data Reduction Pipeline for FIFI-LS, the MIR Integral
  Field Spectrograph for SOFIA}, ed. R.~Pizzo, ASP Conference Series, in press

\bibitem[{{Vogel} {et~al.}(1995){Vogel}, {Weymann}, {Rauch}, \&
  {Hamilton}}]{Vogel1995}
{Vogel}, S.~N., {Weymann}, R., {Rauch}, M., \& {Hamilton}, T. 1995, \apj, 441,
  162, \dodoi{10.1086/175346}

\bibitem[{{Wang} {et~al.}(2010){Wang}, {Zhang}, \& {Fan}}]{Wang2010}
{Wang}, J., {Zhang}, J.-S., \& {Fan}, J.-H. 2010, Research in Astronomy and
  Astrophysics, 10, 915, \dodoi{10.1088/1674-4527/10/9/005}

\bibitem[{{Werner} {et~al.}(2004){Werner}, {Roellig}, {Low}, {Rieke}, {Rieke},
  {Hoffmann}, {Young}, {Houck}, {Brandl}, {Fazio}, {Hora}, {Gehrz}, {Helou},
  {Soifer}, {Stauffer}, {Keene}, {Eisenhardt}, {Gallagher}, {Gautier}, {Irace},
  {Lawrence}, {Simmons}, {Van Cleve}, {Jura}, {Wright}, \&
  {Cruikshank}}]{2004ApJS..154....1W}
{Werner}, M.~W., {Roellig}, T.~L., {Low}, F.~J., {et~al.} 2004, \apjs, 154, 1,
  \dodoi{10.1086/422992}

\bibitem[{{Winkler}(1992)}]{Winkler1992}
{Winkler}, H. 1992, \mnras, 257, 677, \dodoi{10.1093/mnras/257.4.677}

\bibitem[{{Wolfire} {et~al.}(2010){Wolfire}, {Hollenbach}, \&
  {McKee}}]{Wolfire2010}
{Wolfire}, M.~G., {Hollenbach}, D., \& {McKee}, C.~F. 2010, \apj, 716, 1191,
  \dodoi{10.1088/0004-637X/716/2/1191}

\bibitem[{{Zaritsky} {et~al.}(1997){Zaritsky}, {Smith}, {Frenk}, \&
  {White}}]{Zaritsky1997}
{Zaritsky}, D., {Smith}, R., {Frenk}, C., \& {White}, S. D.~M. 1997, \apj, 478,
  39, \dodoi{10.1086/303784}

\end{thebibliography}
\bibliographystyle{aasjournal}


\end{document}